\documentclass[aip,amsmath,amssymb,reprint,twocolumn,jcp,nofootinbib]{revtex4-2}

\usepackage{multirow}
\usepackage{placeins}
\usepackage{graphicx}
\usepackage{amssymb}
\usepackage{mathtools}
\usepackage{xcolor}
\usepackage{booktabs}
\usepackage{siunitx}
\usepackage{comment}
\usepackage{subcaption}
\usepackage[normalem]{ulem}

\newcommand*{\cor}{\text{c}}
\newcommand*{\cond}{\text{cond}}

\newcommand*{\HL}{\text{HL}}
\newcommand*{\Hartree}{\text{H}}
\newcommand*{\kin}{\text{kin}}
\newcommand*{\resp}{\text{resp}}
\newcommand*{\ron}{n}
\newcommand*{\xc}{\text{xc}}
\newcommand*{\xcHole}{\text{xc-hole}}

\DeclarePairedDelimiter{\abs}{\lvert}{\rvert}

\DeclarePairedDelimiter{\norm}{\lVert}{\rVert}
\DeclarePairedDelimiterX\braket[2]{\langle}{\rangle}{#1\delimsize\vert#2}

\newcommand{\ud}{\mathrm{d}}
\newcommand{\br}{\mathbf{r}}
\newcommand*{\bx}{\mathbf{x}}
\newcommand*{\isDefinedAs}{\coloneqq}

\usepackage{hyperref}
\DeclareUnicodeCharacter{2212}{-}

\begin{document}

\title{Secondary kinetic peak in the Kohn-Sham potential and its connection to the response step}
\author{Sara Giarrusso}
\affiliation{Department of Chemistry and Biochemistry, University of California Merced, 5200 North Lake Rd. Merced, CA 95343, USA}
\author{Roeland Neugarten}
\affiliation{Department of Theoretical Chemistry and Amsterdam Center for Multiscale Modeling, FEW, Vrije Universiteit, De Boelelaan 1083, 1081HV Amsterdam, The Netherlands}
\author{Evert Jan Baerends}
\affiliation{Department of Theoretical Chemistry and Amsterdam Center for Multiscale Modeling, FEW, Vrije Universiteit, De Boelelaan 1083, 1081HV Amsterdam, The Netherlands}
\author{Klaas J. H. Giesbertz}
\affiliation{Department of Theoretical Chemistry and Amsterdam Center for Multiscale Modeling, FEW, Vrije Universiteit, De Boelelaan 1083, 1081HV Amsterdam, The Netherlands}

\date{\today}

\begin{abstract}
We consider a prototypical 1D model Hamiltonian for a stretched heteronuclear molecule and construct individual components of the corresponding KS potential, namely: the kinetic, the $N-1$, and the conditional potentials.
These components show very special features, such as peaks and steps, in regions where the density is drastically low. Some of these features are quite well known, whereas others, such as a secondary peak in the kinetic potential or a second bump in the conditional potential, are less or not known at all.
We discuss these features building on the analytical model treated in \href{https://pubs.acs.org/doi/10.1021/acs.jctc.8b00386}{J.~Chem.\ Theory Comput.~\textbf{14}, 4151 (2018)}.
In particular, we provide an explanation for the underlying mechanism which determines the appearance of both peaks in the kinetic potential and elucidate why these peaks delineate the region over which the plateau structure, due to the $N-1$ potential, stretches.
We assess the validity of the Heitler--London \textit{Ansatz} at large but finite internuclear distance, showing that, if optimal orbitals are used, this model is an excellent approximation to the exact wavefunction. 
Notably, we find that the second natural orbital 
presents an extra node very far out on the side of the more electronegative atom. 
\end{abstract}

\maketitle

\section{Introduction}

Kohn--Sham Density Functional Theory (KS-DFT) incorporates the effects of electrons interacting with one-another through an auxiliary system which is formally non-interacting.
The external potential of such auxiliary system -- the KS potential -- contains the external potential of the original interacting system plus the so-called  Hartree-exchange-correlation (Hartree-XC) potential, which effectively replaces the two-body interaction operator.  
Via this mapping, any ground state property of the interacting system can be calculated from knowledge of its Hartree-XC potential, by solving a number of one-body Schrödinger equations. Unfortunately, only its mean-field part called the Hartree potential, $v_{\text{H}}$, is known universally, and the remainder, called the XC potential, $v_\xc$, must be approximated by models. 

LDA and GGA functionals offer an adequate approximation to a component of the XC potential known as the exchange-correlation hole potential. This term embodies the electron correlation effect which is needed to obtain the correct interaction energy of the physical system, and therefore it has a direct role in the energetics of quantum systems. However, there are other components of the XC potential that are more elusive: they often have a minor energetic role (if any) but are nonetheless necessary to generate a non-interacting density that matches the interacting one and are related to different correlation effects.

These other components typically show quite dramatic features, such as ``peaks" and ``steps". As an example, features of this kind arise in the XC potential of molecular species in which a covalent bond is stretched. Despite the enormous success of a great many density functional approximations
, an accurate modelling of the XC potential in these cases is still lacking and dissociation processes still represent a major challenge for KS-DFT.

\begin{figure}[t]
\includegraphics[width=\columnwidth]{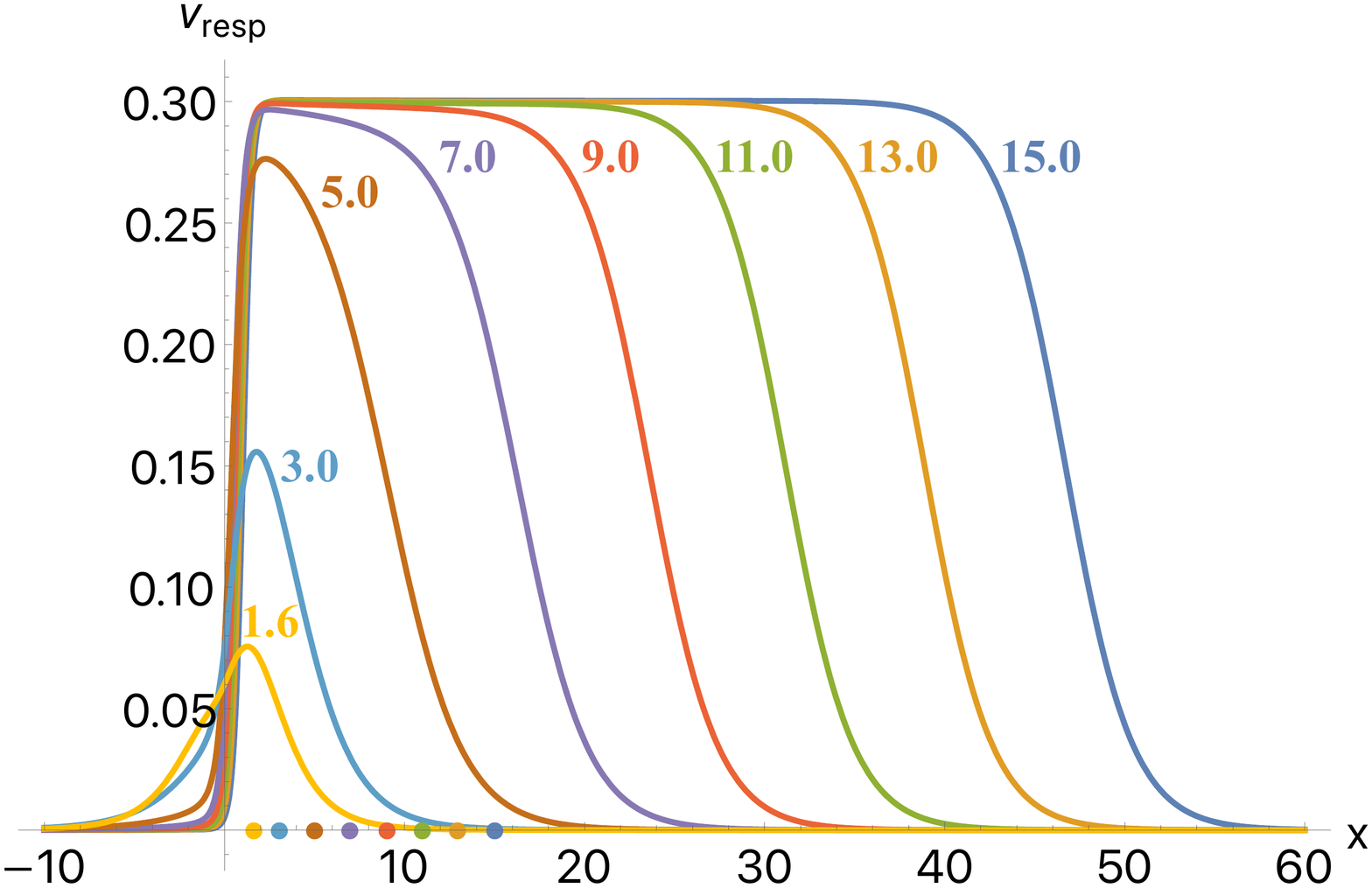}
\caption{The response potential defined in~\eqref{eq:vNm1} at different values of the internuclear distance, $R$ in atomic units indicated next to the plot lines. The location of the right nucleus A at each internuclear distance at $\frac{R}{2}$ is marked by a dot.}
\label{fig:resp_het}
\end{figure}

Occurrences of peaks and steps have been investigated in several theoretical studies, where the exact potential has been computed accurately for some simple\slash{}model systems. 
Pioneering work in this direction has been carried out in the nineties.\cite{BuiBaeSni-PRA-89, GriLeeBae-JCP-94, LeeGriBae-ZPD-95, GriBae-PRA-96, GriLeeBae-JCP-96, BaeGri-JPCA-97} In these papers, based on the theory of the conditional amplitude\cite{Hun-IJQC-75-1, Hun-IJQC-75-2}
 an exact expression for the Hartree-XC potential was derived in terms of physically transparent terms. Apart from the XC hole potential $v_{\xcHole}$, they identified the so-called response potential $v_{\resp}$~\eqref{eq:vrespNm1} as well as the correlation kinetic potential $v_{\cor,\kin}$~\eqref{eq:vckin}. The definition of these potentials will be  
 recapitulated in Sec.~\ref{sec:CAdec}.

The best known special feature of the KS potential is a positive shift (plateau) in the XC potential that builds up over the site of the more electronegative atom in the case of a stretched heteronuclear diatomic system. This has been deduced as early as the mid ’80s~\cite{AlmBar-INC-85, Per-INC-85} with the qualitative argument of necessary ``equalization'' of the different ionization potentials of the open-shell fragments, in order to have equal distribution of the two electrons over the fragments. This feature has received further interest in the literature.\cite{TemMarMai-JCTC-09,HelTokRub-JCP-09,Maitra2012KSpotTDDFT,Hodgson2016,Hodgson2017,GiaVucGor-JCTC-18, GiaGorGie-EPJB-18, GiaGor-JPCA-20,Kraisler2021} In Ref.~\onlinecite{GriBae-PRA-96}, it was demonstrated that such step structure (with height $\Delta I = I_A - I_B$ if A has higher ionization potential) is due to the response potential $v_{\resp}$. This is illustrated in Fig.~\ref{fig:resp_het} where the exact $v_{\resp}$ is shown from numerically very precise solutions of a diatomic model system described in Sec.~\ref{sec:COMPdet}. The plateau character of the potential is very clearly demonstrated. It is also to be noted that the width of the plateau is accurately rendered by our calculations and is dependent on the distance $R$ between the nuclei. The precise extent of the plateau is however considerably larger than the bond distance $R$ (the nuclei are at $\pm R/2$) and calls for an explanation.

\begin{figure}[t]
\includegraphics[width=\columnwidth]{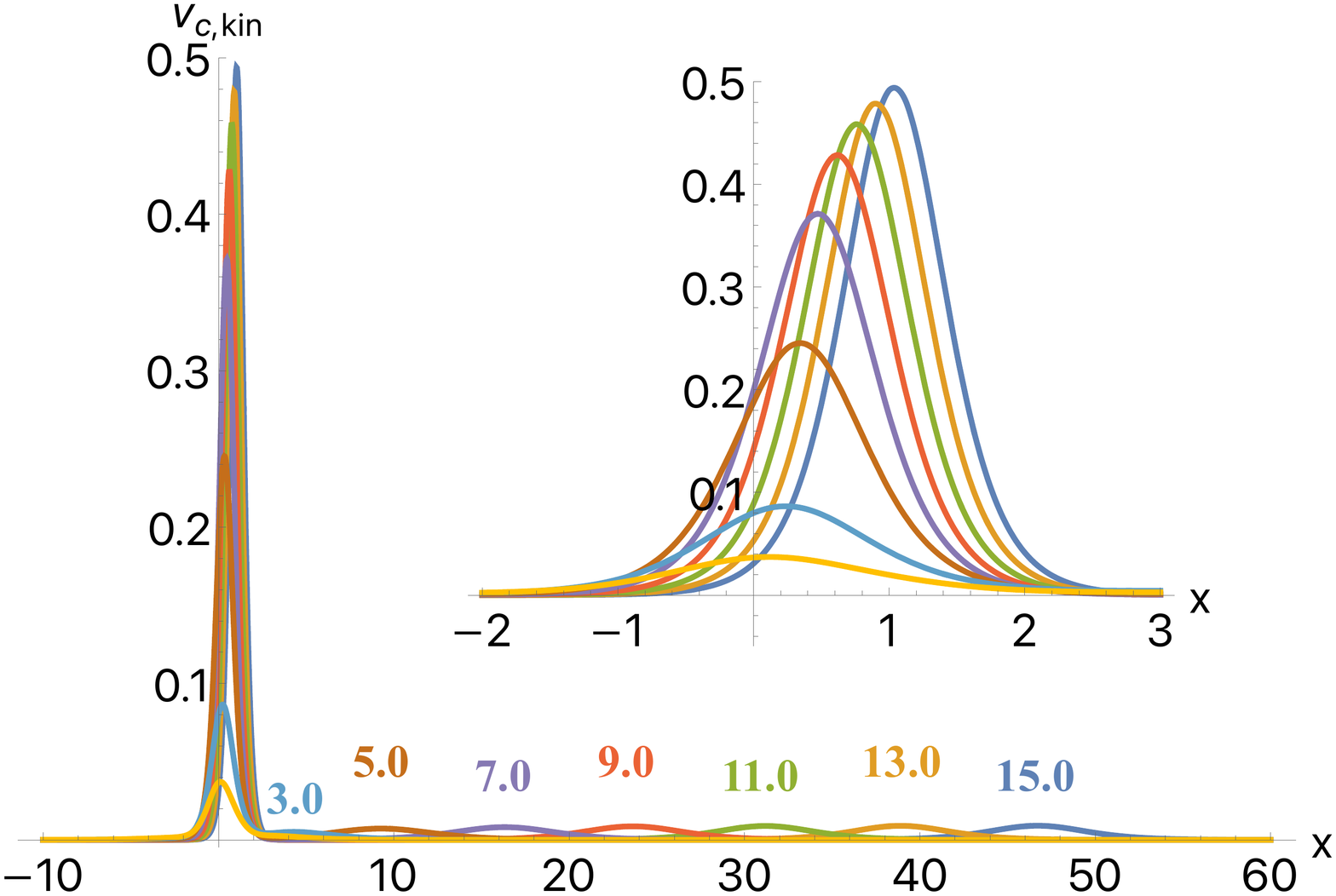} \\
\includegraphics[width=\columnwidth]{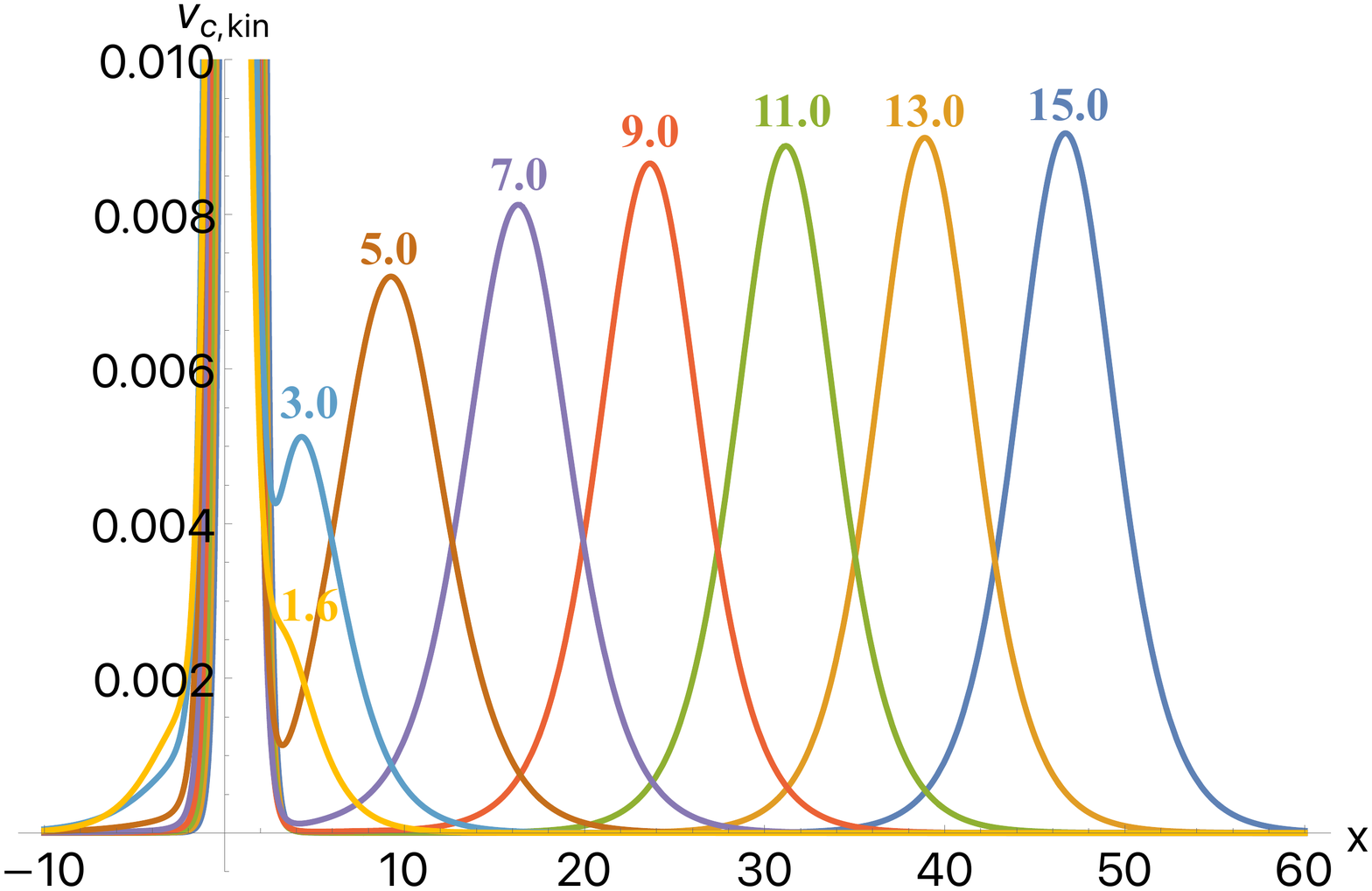}
\caption{Upper panel: the (correlation) kinetic potential~\eqref{eq:vkin}, $v_{\cor, \kin}$, in full scale and zoomed on the x-axis around the origin to show the primary peak (inset); lower panel: $v_{\cor, \kin}$ zoomed in in the range where the secondary peak shows up.}
\label{fig:2nd_kin_peak_het}
\end{figure}

Another feature which is known to appear in the XC potential of stretched diatomic systems is a peak in the bond midpoint region.\cite{BuiBaeSni-PRA-89, GriBae-PRA-96, GriBae-TCA-97, BaeGri-JPCA-97, HelTokRub-JCP-09, TemMarMai-JCTC-09} It originates from  the correlation kinetic potential, $v_{\cor, \kin}$~\eqref{eq:vckin}. It is displayed in the upper panel of Fig.~\ref{fig:2nd_kin_peak_het}. 
Recently, a second peak of lower intensity in the same $v_{\cor, \kin}$ potential has been identified\cite{GiaVucGor-JCTC-18} using a 1D model which becomes ``exact” at infinite internuclear separation and it has more recently been reported also in a similar model from numerically solving a 1D Schrödinger equation.\cite{KocJakEliAxe-arxiv-21}  Such secondary peak is also shown in Fig.~\ref{fig:2nd_kin_peak_het}: it is much lower than the bond midpoint peak and barely visible in the upper panel, so it is blown up in the lower panel. It is not a spurious feature, but it is very distinct and its position is obviously related to the extent of the plateau. The plateau actually seems to be straddled by the high bond midpoint peak and the low second kinetic peak, which is another intriguing feature of the KS potential that is to be explained.

The paper is structured as follows. In Sec.~\ref{sec:CAdec} we review the decomposition of the KS potential and define the various terms, notably the response potential and the kinetic potential. Then in Sec.~\ref{sec:HLreview} explanations of the structure of these potentials is given with the help of the Heitler--London (HL) wavefunction using the the treatment of Ref.~\onlinecite{GiaVucGor-JCTC-18}. Despite its simplicity, such an asymptotically correct model is useful since it affords the derivation of analytical results and highlights the origin of the special features of the potential. It provides a graphic illustration of what we have named  the ``jumping" of the conditional amplitude (see Fig.~\ref{fig:CAjump}). This is the mechanism through which the appearance of the bond midpoint peak has been explained\cite{BuiBaeSni-PRA-89} and which now provides an explanation of the secondary peak on the far side of the more electronegative atom. This represents one of the core results of this work (Sec.~\ref{sec:HLreview}).
The explanations on the basis of the simple HL model should be verified by actual calculations. To this end we introduce in Sec.~\ref{sec:modelHam} a model Hamiltonian that allows very accurate solution. Extreme numerical precision (see Appendix~\ref{sec:numdetails}) is required since, as can be seen in Figs~\ref{fig:resp_het} and~\ref{fig:2nd_kin_peak_het}, the plateau extends very far out into regions of extremely low density and the second kinetic peak occurs equally far out. Section~\ref{sec:Results} presents our precise numerical results, validating the deductions from the HL model and providing additional explanation.
In Sec.~\ref{sec:Discussion}, we report a novel feature of the conditional potential directly related to the jumping of the conditional amplitude and discuss the concept of atomic regions in a strecthed molecule.
 Finally, Sec.~\ref{sec:conclusions} contains some conclusive thoughts that generalise our results to $N$-electron systems.

\section{Decomposition of the KS potential}\label{sec:CAdec}
According to the work of Buijse \textit{et al}.~\cite{BuiBaeSni-PRA-89} the Hartree-XC potential, $v_\text{Hxc}$, can be exactly decomposed as
\begin{align}\label{eq:KSpotdec}
v_\text{Hxc}(\br_1) ={}& v_{\kin}(\br_1) - v_{s,\kin}(\br_1) \notag \\
&{}+ v_{N-1}(\br_1) - v_{s,N-1}(\br_1) + v_{\cond} (\br_1).
\end{align}
Each of the contributions on the r.h.s.\ are defined in terms of the so-called \emph{conditional (probability) amplitude} \cite{Hun-IJQC-75-1, Hun-IJQC-75-2}
\begin{equation}\label{eq:CAdef}
\Phi (\sigma_1, \bx_2,\dotsc, \bx_N;\br_1) 
\isDefinedAs \frac{\Psi (\bx_1,\bx_2,\dotsc, \bx_N) \, \sqrt{N}}{\sqrt{\ron (\br_1)}}
\end{equation}
with $\Psi (\bx_1,\dotsc, \bx_N)$ and $\ron (\br_1)$ respectively the many-body ground state (GS) wavefunction and the GS density of the $N$-electron Hamiltonian under study and with $\bx_1 \isDefinedAs \br_1\sigma $ and $\bx_i \isDefinedAs \br_i \sigma_i$, i.e.\ the product of the spatial and spin coordinates of the particles.
They are
\begin{align}\label{eq:vkin}
v_{\kin}(\br_1) &\isDefinedAs 
\frac{1}{2}\int \abs{\nabla_{\br_1}\Phi}^2\, \ud \sigma_1 \ud \bx_2 \dotsi \ud \bx_N, \\
\label{eq:vNm1}
v_{N-1}(\br_1) &\isDefinedAs 
\int \Phi^* \hat{H}^{N-1}\Phi\, \ud \sigma_1 \ud \bx_2 \dotsi \ud \bx_N -E_0^{N-1},
\end{align}
with $ \hat{H}^{N-1}$ and $E_0^{N-1}$ respectively the Hamiltonian and the GS energy of the ionised system, and
\begin{equation}\label{eq:vcond}
v_{\cond}(\br_1) \isDefinedAs \int \sum_{i=2}^N v_{ee}(\br_1 -\br_i)\abs{\Phi}^2 \, \ud \sigma_1 \ud \bx_2 \dotsi \ud \bx_N
\end{equation}
with $v_{ee}$ being any interaction function.

Finally, $v_{s,\kin}$ and $v_{s,N-1}$ are formally defined as in~\eqref{eq:vkin} and~\eqref{eq:vcond} but having the KS conditional amplitude, $\Phi_s$, in place of the interacting one, $\Phi$. The KS conditional amplitude $\Phi_s$ comes from~\eqref{eq:CAdef} replacing $\Psi$ with the KS wavefunction. 
 
The physical interpretation of the potentials just introduced hinges upon the physical meaning of the conditional amplitude~\eqref{eq:CAdef}. The modulus square of the conditional amplitude represents the probability of finding the $N-1$ particles at coordinates $\{ \bx_2, \dots, \bx_N\}$ given that the reference electron is in position $\br_1$. Because the probability of finding the $N-1$ electrons anywhere in space is one regardless of where the reference electron is positioned, then
\begin{equation}\label{eq:Phinorm}
\int \abs{\Phi (\sigma_1, \bx_2,\dotsc, \bx_N;\br)}^2 \ud \sigma_1 \ud \bx_2 \dotsi \ud \bx_N = 1 \quad \forall \,\br_1,
\end{equation}
a condition which is sometimes referred to as \emph{partial normalisation condition}.
 In this sense, $\Phi $ is a sort of $(N-1)$-electron wavefunction that describes how the electronic cloud of $N-1$ electrons is influenced by the presence of a reference electron.
 
The term with the easiest physical interpretation is the conditional potential~\eqref{eq:vcond} which corresponds to the electrostatic potential of the conditional density, i.e.\ $v_{\cond}(\br) = \int \ron_\cond(\br'; \br)\, v_{ee}(\br -\br')\,\ud \br'$, with $\ron_\cond(\br';\br)=\frac{P_2 (\br, \br')}{\ron(\br)} $ and $P_2(\br, \br')$ the pair-density.
Furthermore, this potential is usually seen as made up of two contributions:  the Hartree potential, $v_\text{H}$, plus the exchange-correlation hole potential, $v_\xcHole$. This latter can be written as
\begin{equation}
v_\xcHole (\br) = \int \ron (\br') \,g_{\xc}(\br, \br')\, v_{ee}(\br -\br') \,\ud \br'
\end{equation}
with $g_{\xc}(\br, \br') \isDefinedAs \frac{P_2 (\br, \br')}{\ron(\br) \ron (\br')} -1$. Thus, $v_\xcHole$ represents the electrostatic potential of the exchange-correlation hole, $h_{\xc}(\br, \br') =\ron (\br') \,g_{\xc}(\br, \br')$. 

The kinetic potential~\eqref{eq:vkin} contains the gradient of the conditional amplitude w.r.t.\ the position of the reference electron; therefore it is expected to peak in regions where the conditional amplitude changes character rapidly, in ``sensitive" regions. As we have mentioned in the introduction, this is known to happen near the bond midpoint region of a stretched covalent bond.

Lastly, the $v_{N-1}$ potential in~\eqref{eq:vNm1} represents a sort of local energy of the $N-1$ system
with respect to some baseline energy which in~\eqref{eq:vNm1} is given by the energy of the singly ionised system. Also this term has been shown to play a crucial role when a covalent bond between heteronuclear fragments is stretched, by building a plateau around the more electronegative fragment.

All these potentials typically go to zero asymptotically. Roughly speaking, this is a consequence of the fact that the conditional amplitude becomes less and less sensitive to the position of the reference electron as its distance from the bulk of the density increases, an argument which breaks down in the presence of nodal planes,\cite{GorGalBae-MP-16, GorBae-EPJB-18} which we do not consider in this work.

Some of the contributions on the r.h.s.\ of~\eqref{eq:KSpotdec} may appear, in the literature, as collected into more compact terms, by introducing the definitions of the correlation kinetic potential, $v_{\cor,\kin} $
\begin{equation}
v_{\cor,\kin} \isDefinedAs  v_{\kin}-v_{s, \kin}\label{eq:vckin}, 
\end{equation}
which delivers the correlation kinetic energy  when multiplied by the density and integrated over space, i.e.\ $T_{\cor}[\ron]=\int \left( v_{\cor,\kin}(\br) \ron(\br) \right) \ud \br $, and of the response potential
\begin{equation}
v_{\resp} \isDefinedAs  v_{N-1}-v_{s, N-1} \label{eq:vrespNm1}.
\end{equation}
As a final note, we point out that, in the special case of two-electron singlets, the KS kinetic and $N-1$ potentials are zero.
This is readily seen for the former since the KS conditional amplitude reduces to a one-body function equal to the square root of the density divided by two, i.e.\ $\sum_{\sigma, \sigma_2} \Phi_s (\sigma, \bx_2; \br) =\sqrt{\frac{\ron (\br_2)}{2}}$, losing the parametric dependence on the position of the reference electron.
As for the response potential, we have 
\begin{align}
v_{s, N-1} & = \!\int\! \sqrt{\frac{\ron(\br_2)}{2}} \left( - \frac{\nabla_{\br_2}^2}{2}+v_s (\br_2)\right) \sqrt{\frac{\ron(\br_2)}{2}} \, \ud \br_2 - \epsilon_\text{H} \notag \\
& = \frac{\epsilon_\text{H}}{2} \int \ron(\br_2)\,\ud\br_2- \epsilon_\text{H}=0.
\end{align}
Therefore, in these cases, we have
\begin{subequations}
\begin{align}
v_{c,\kin} & \equiv  v_{\kin} \label{eq:vckinvkin} \\
v_{\resp} & \equiv  v_{N-1}\label{eq:vrespvnm1}. 
\end{align}
\end{subequations}

\section{Features of the kinetic potential from a simple Heitler--London \textit{Ansatz}}\label{sec:HLreview}

In this section we review and extend some analytical results on the kinetic potential for a dissociating AB molecule reported in Ref.~\onlinecite{GiaVucGor-JCTC-18}. Furthermore, we provide an explanation for the appearance of the secondary peak based on the structure of the conditional amplitude.

\subsection{Analytical expression for the kinetic potential}
In Ref.~\onlinecite{GiaVucGor-JCTC-18},  using the Heitler--London  (HL) \textit{Ansatz}, which is an accurate model for a two-electron molecular system in the dissociation limit,
\begin{equation}\label{eq:HLWF}
\Psi_{\HL}(x_1, x_2)
= \frac{ \phi_A(x_1) \phi_B(x_2) + \phi_B(x_1)\phi_A(x_2)}{\sqrt{2\left(1+S^2 \right)}},
\end{equation}
(with $S$ the overlap $\braket{\phi_A}{\phi_B}$) the kinetic potential has been computed analytically [\onlinecite[see Eq.~(81) in Ref.][]{GiaVucGor-JCTC-18}]. 
In particular, a simple exponential basis for the two fragment orbitals
\begin{equation}\label{eq:expbases}
\phi_{A(B)}= \sqrt{\frac{a (b)}{2}}\, e^{-\frac{a (b)}{2} \abs{x \, -\!(+) \frac{R}{2}}},
\end{equation}
with exponents $a>b$, to mimic the effect of different nuclei separated by a distance $R$, has revealed the presence of a secondary peak in the kinetic potential besides the very well-known primary one.\cite{BuiBaeSni-PRA-89, GriBae-PRA-96, BaeGri-JPCA-97, TemMarMai-JCTC-09, HelTokRub-JCP-09}
%
%
Furthermore, with this model, the locations of the two peaks and their maximal values as functions of the ionisation potential of each fragment have been derived and tabulated, \onlinecite[see Table~I in Ref.][]{GiaVucGor-JCTC-18} and~\eqref{eq:xeq} and~\eqref{eq:vkinmax} below. (Ionisation potentials are related to the exponents as $I_{\alpha}=\alpha^2/8$, $\alpha=a,b$.)
Such analytical expressions were obtained by considering the overlap integral $S=\braket{\phi_A}{\phi_B}$ as exactly zero, and using the corresponding conditional amplitude. 
However, using the more general expression
\begin{equation}
\Phi_\text{HL} = \frac{\left( \phi_A(x_1) \phi_B(x_2) + \phi_B(x_1)\phi_A(x_2) \right)}{\sqrt{\phi_A(x_1)^2+\phi_B(x_1)^2+2\, S\phi_A(x_1)\phi_B(x_1)}},
\end{equation}
one obtains an expression for the kinetic potential which is essentially the same from a qualitative point of view, namely
\begin{equation}\label{eq:vkinwithoverlap}
v_\kin^\text{HL}= \frac{\left( 1-S^2\right)}{2}\frac{\left(\phi_B\,\phi_A' - \phi_A\,\phi_B'\right)^2}{\left( \phi_A^2 + \phi_B^2  +2\,S \phi_A \phi_B \right)^2}.
\end{equation}
The expressions for the location of the peaks and their height, using the simple exponential bases ($a>b$), become
\begin{align}
x_{\text{peak}}^{\pm} &= \frac{(a \pm b)R \pm 2\ln\bigl(\frac{a}{b}\bigr) }{2(a \mp b)} \label{eq:xeq} \\
v_\kin\bigl(x_{\text{peak}}^{\pm}\bigr) &= \frac{1}{8}\left(\frac{a \mp b}{2}\right)^{\!2} \label{eq:vkinmax}
\frac{1 - \tilde{S}}{1 + \tilde{S}} ,
\end{align}
with $-R/2 < x_{\text{peak}}^{-} < R/2 < x_{\text{peak}}^{+}$ and $\tilde{S}$ is the overlap of the exponential basis functions
\begin{equation}\label{eq:S}
\tilde{S}= \frac{2 \sqrt{a \,b} \left(a \,e^{-\frac{b R}{2}}-b\, e^{-\frac{a R}{2}}\right)}{a^2-b^2}.
\end{equation} 
Apparently the value of the overlap only affects the height of the kinetic peaks (by reducing them) but not their locations. 
We call the peak between the two nuclei, at $x_{\text{peak}}^{-}$, the primary peak (the higher one) and the low peak at $x_{\text{peak}}^{+}$ the secondary peak. Clearly, from~\eqref{eq:xeq} if the system were symmetrical ($a=b$) the position of the first peak would be exactly at $x=0$, as expected, and the second peak would disappear to infinity. In a heterogeneous system ($a>b$), the secondary peak appears well beyond the nucleus to the right, as is evident from writing it as  $x_{\text{peak}}^+= (a-b)^{-1}\left((a+b)R/2+\ln{\frac{a}{b}}\right)$. This agrees with Fig.~\ref{fig:2nd_kin_peak_het} (we use as example $a=2$ and $b=1$, with $A$ to the right). The points $x_{\text{peak}}^-$ and $x_{\text{peak}}^+$ match the location of the inflexion points of $v_{\resp}$ exactly in the HL model with exponential basis functions~\eqref{eq:expbases} and may be taken as the border points of the plateau of Fig.~\ref{fig:resp_het} (see also end of Sec.~\ref{sec:JoCA}). It is then easy to see that the plateau has a width proportional to $R$ and extends considerably beyond the rightmost nucleus at $R/2$,
\begin{equation}\label{eq:plateauwidth}
x_{\text{peak}}^+ - x_{\text{peak}}^- = \frac{2abR+2a\ln{\frac{a}{b}}}{a^2-b^2}.
\end{equation} 
as was noted in connection with Fig.~\ref{fig:resp_het}.

Note that $x_{\text{peak}}^{\pm}$ correspond to where $\phi_A \equiv \phi_B$ when we use the simple exponential basis~\eqref{eq:expbases}. We will discuss the relation of the kinetic peak positions with  the crossings of the localised (atomic) orbitals, and the corresponding nodes in delocalised orbitals, in Sec.~\ref{subsec:crossings}, but first address in the next section the physics behind these peaks as derived from the ``jumping'' behaviour of the conditional amplitude.

\begin{figure*}
\centering
 \begin{tabular}[c]{cc}
 {\begin{subfigure}{0.5\textwidth}
      \includegraphics[scale=0.44]{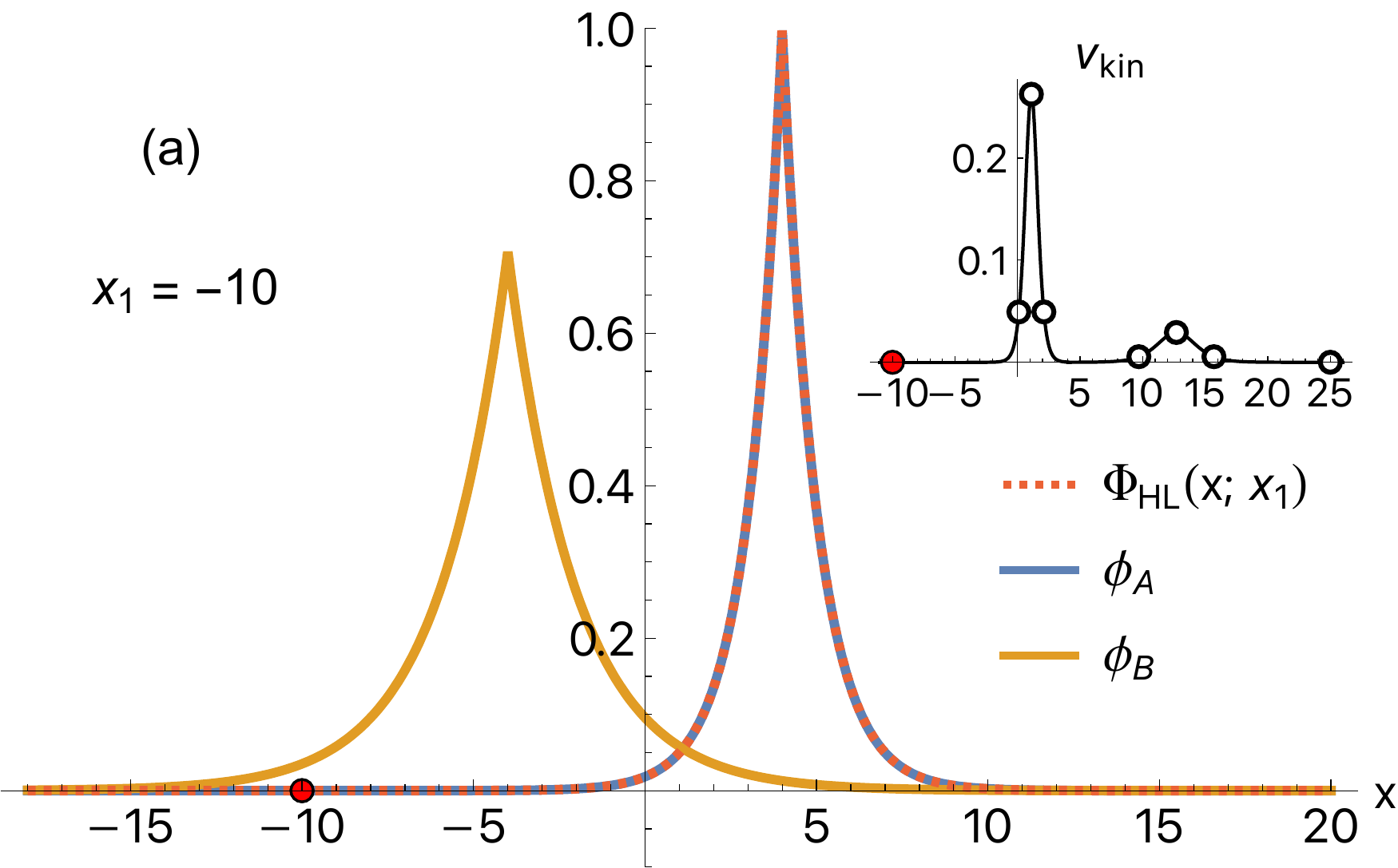}
    \end{subfigure}
} & {\begin{subfigure}{0.5\textwidth}
 \includegraphics[scale=0.44]{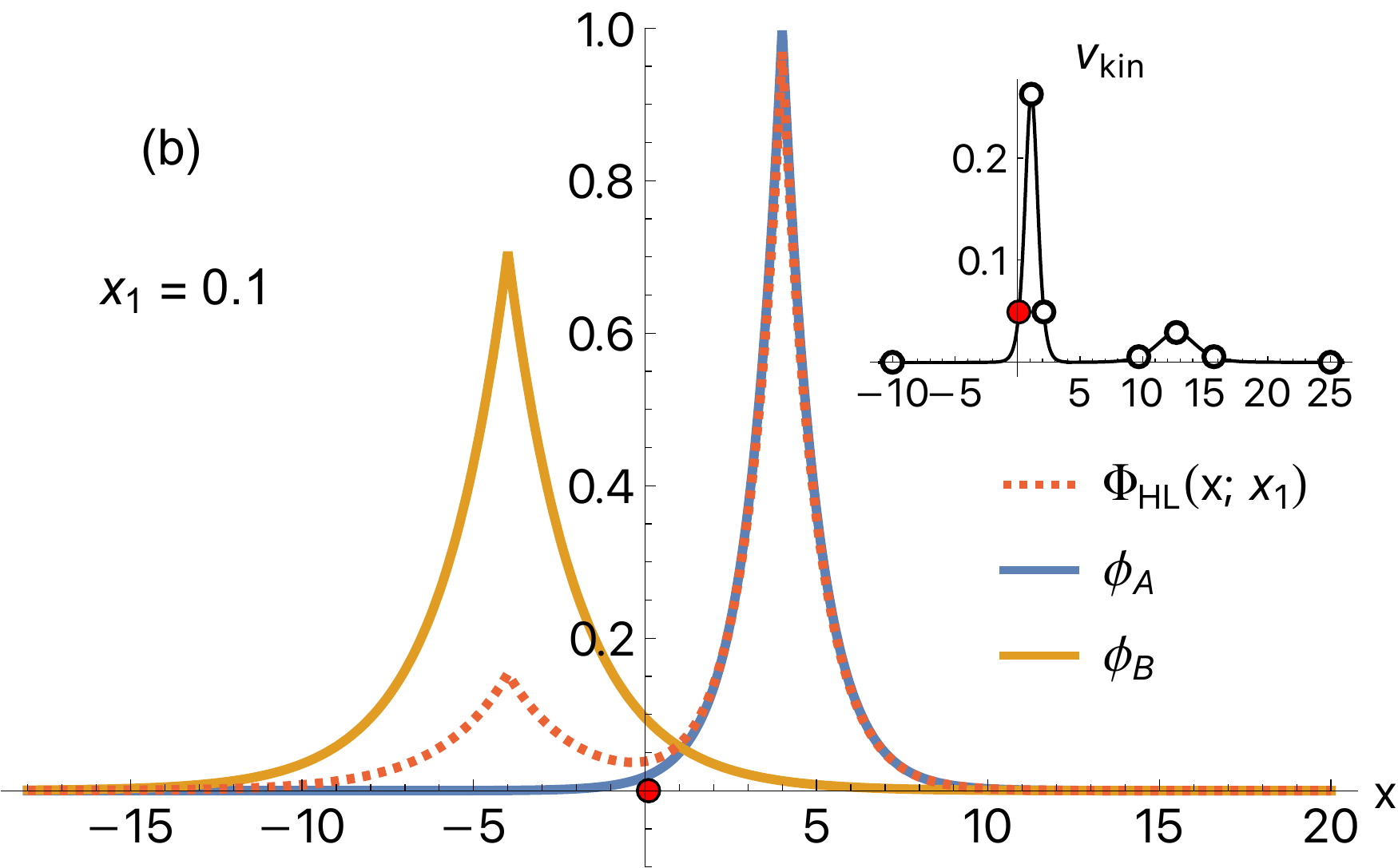}
 \end{subfigure}}\\
  {\begin{subfigure}{0.5\textwidth}
 \includegraphics[scale=0.44]{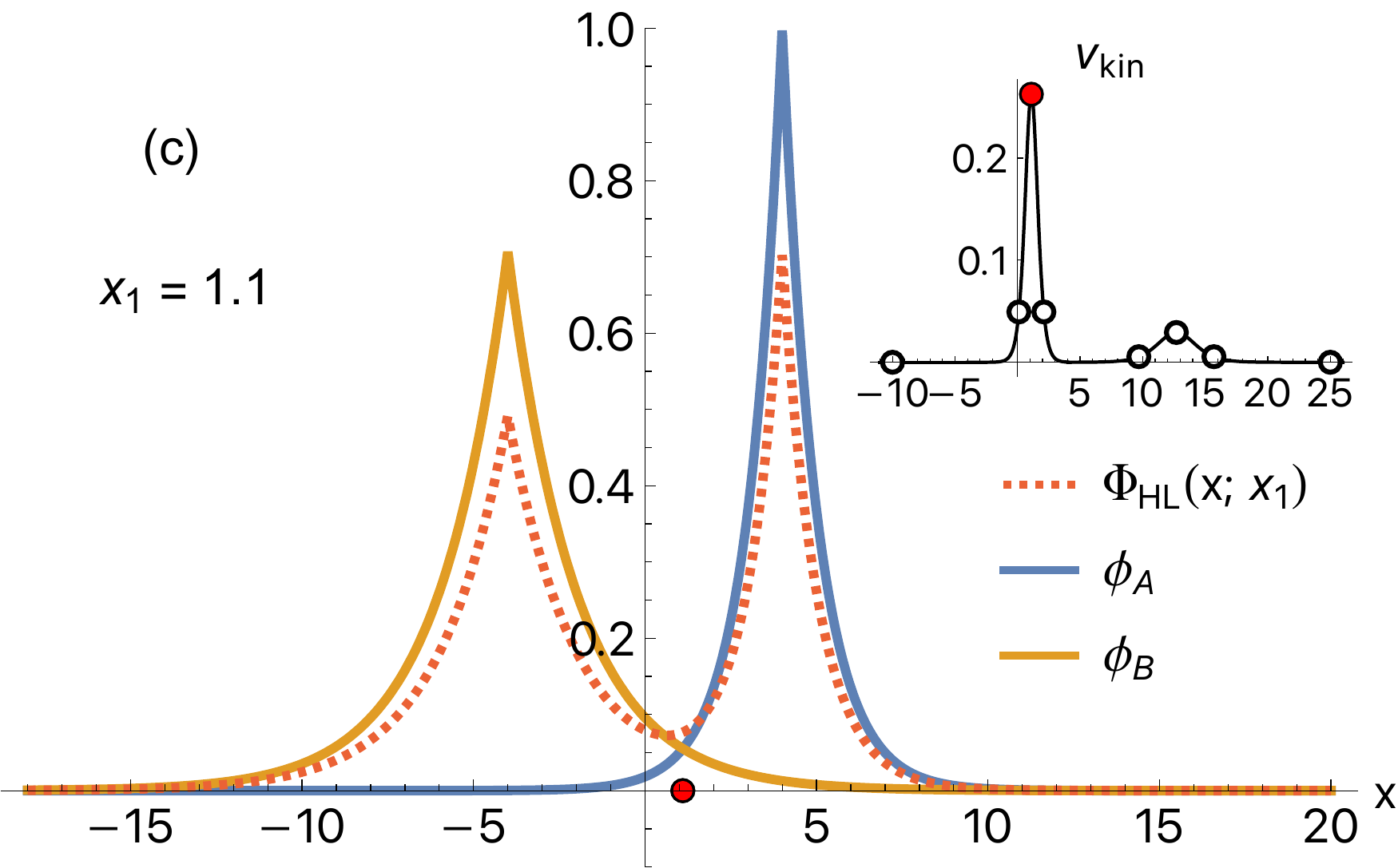}
 \end{subfigure}}
 & {\begin{subfigure}{0.5\textwidth}
 \includegraphics[scale=0.44]{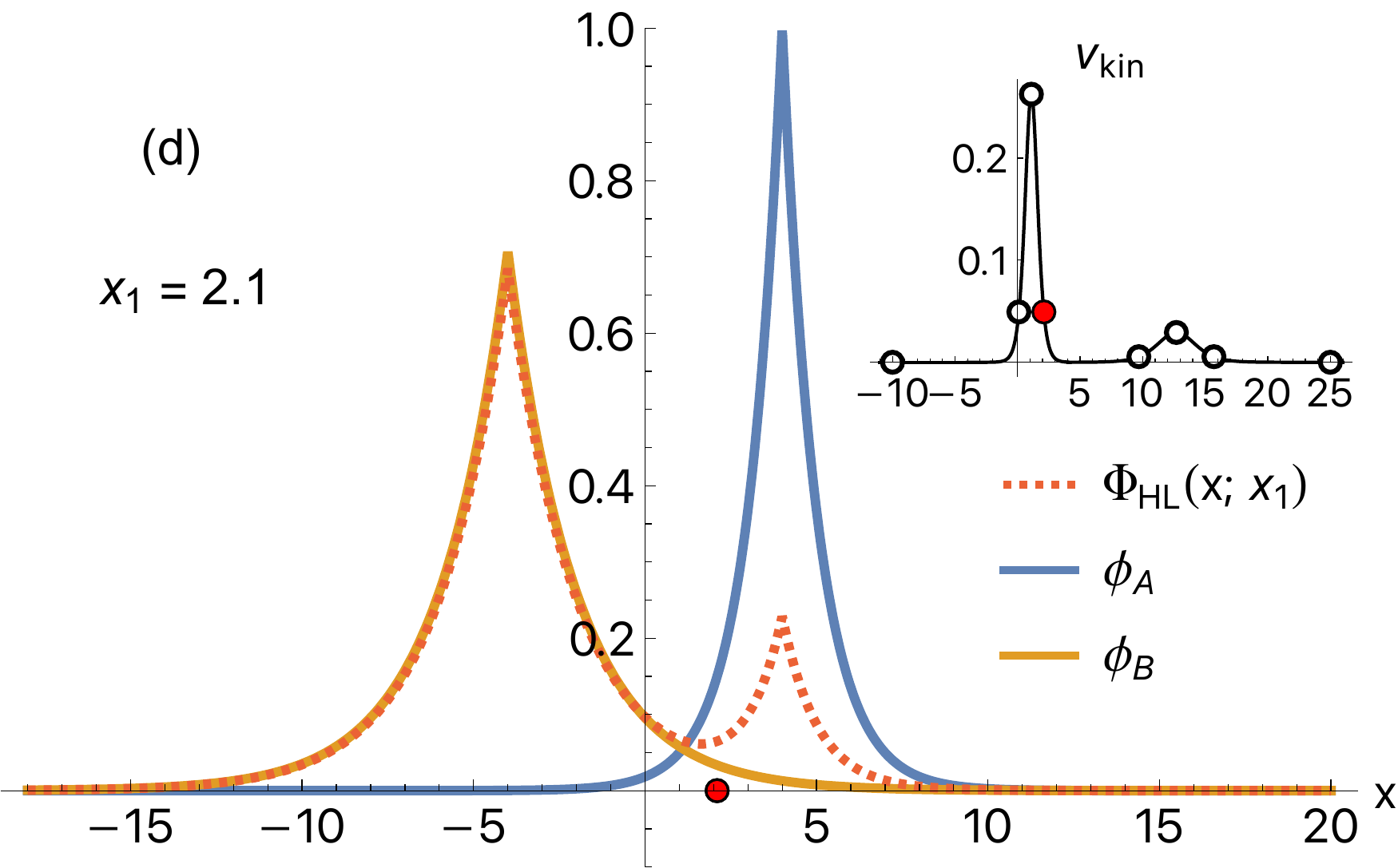}
 \end{subfigure}}\\
 {\begin{subfigure}{0.5\textwidth}
      \includegraphics[scale=0.44]{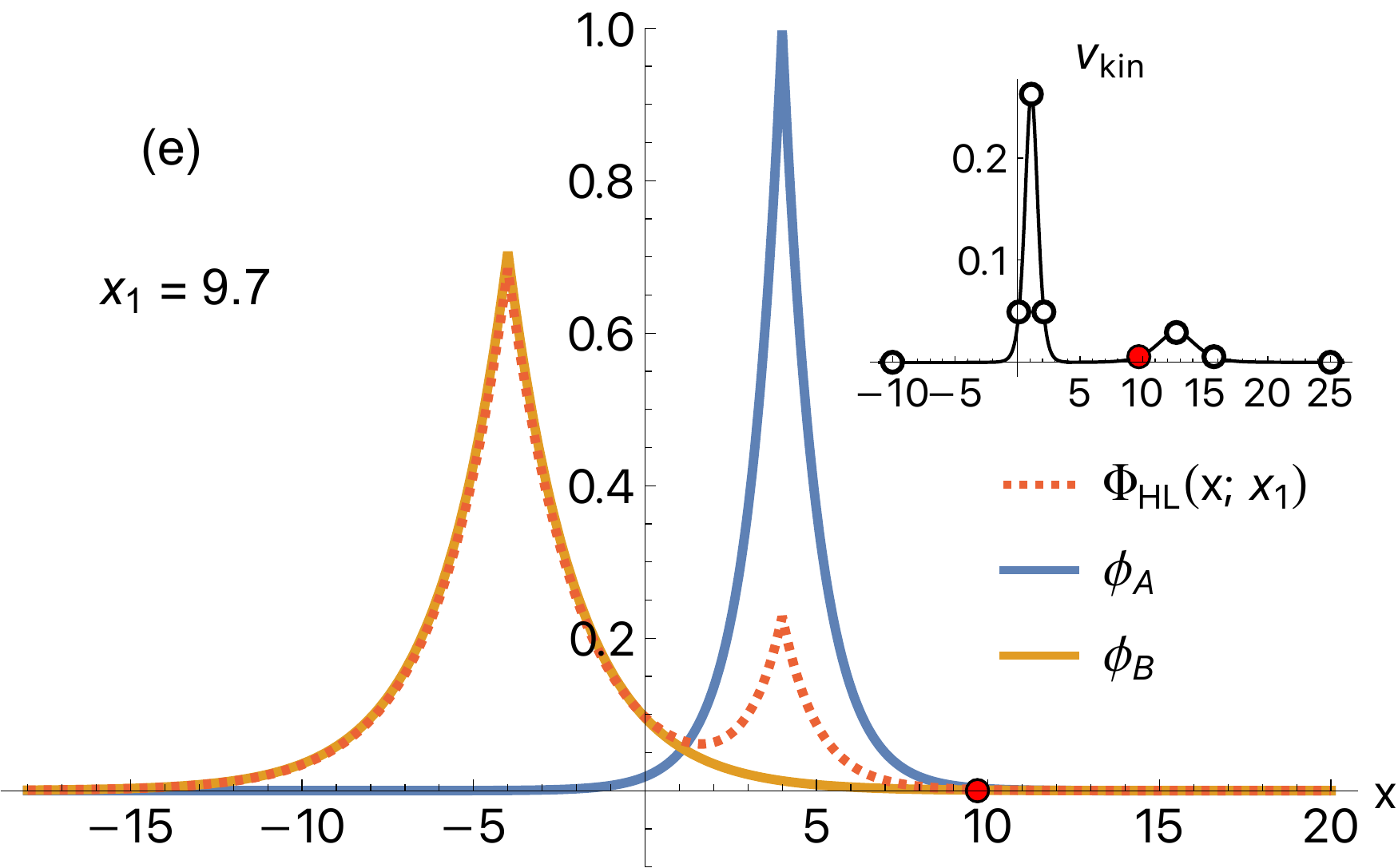}
    \end{subfigure}
} & {\begin{subfigure}{0.5\textwidth}
 \includegraphics[scale=0.44]{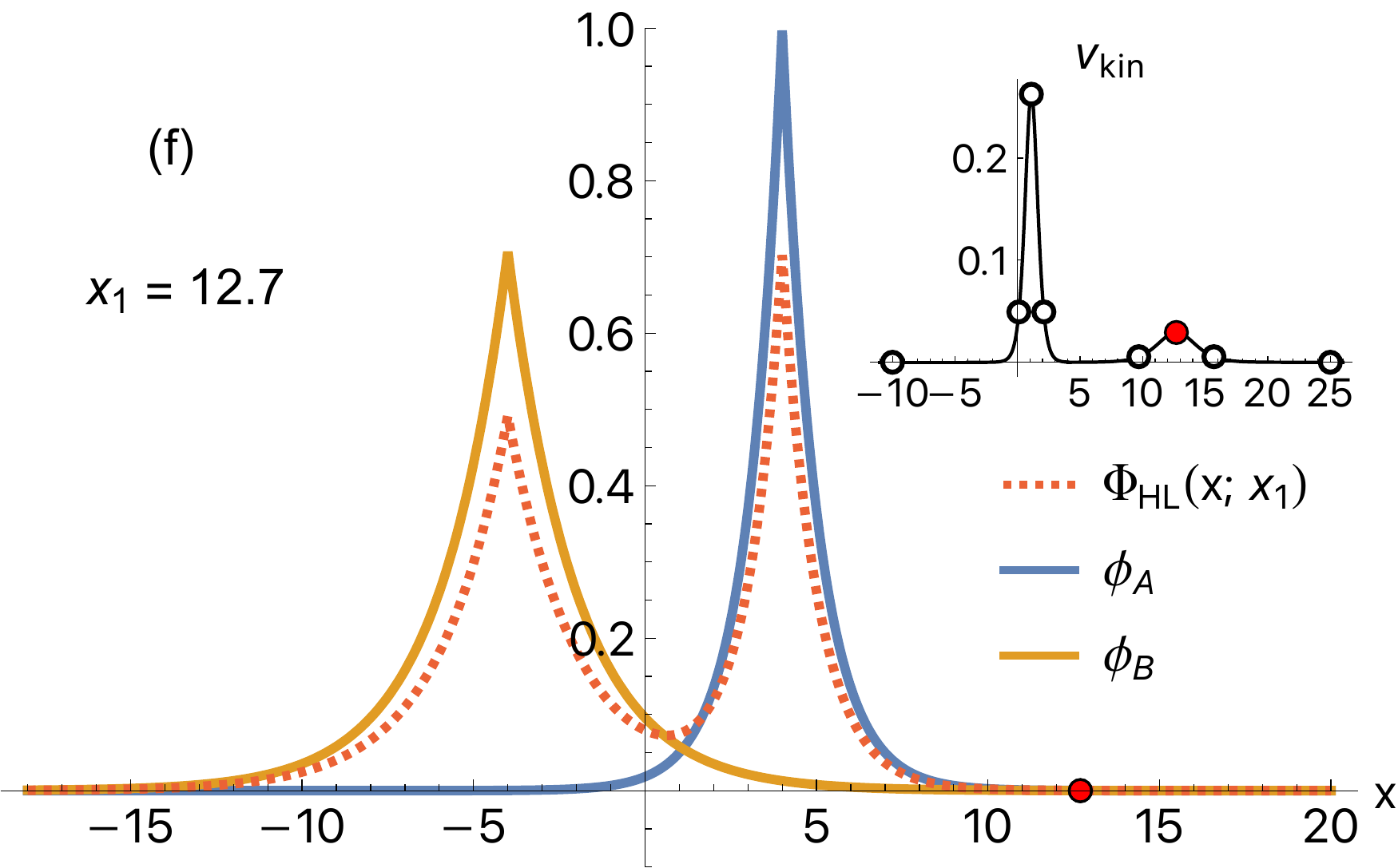}
 \end{subfigure}}\\
  {\begin{subfigure}{0.5\textwidth}
 \includegraphics[scale=0.44]{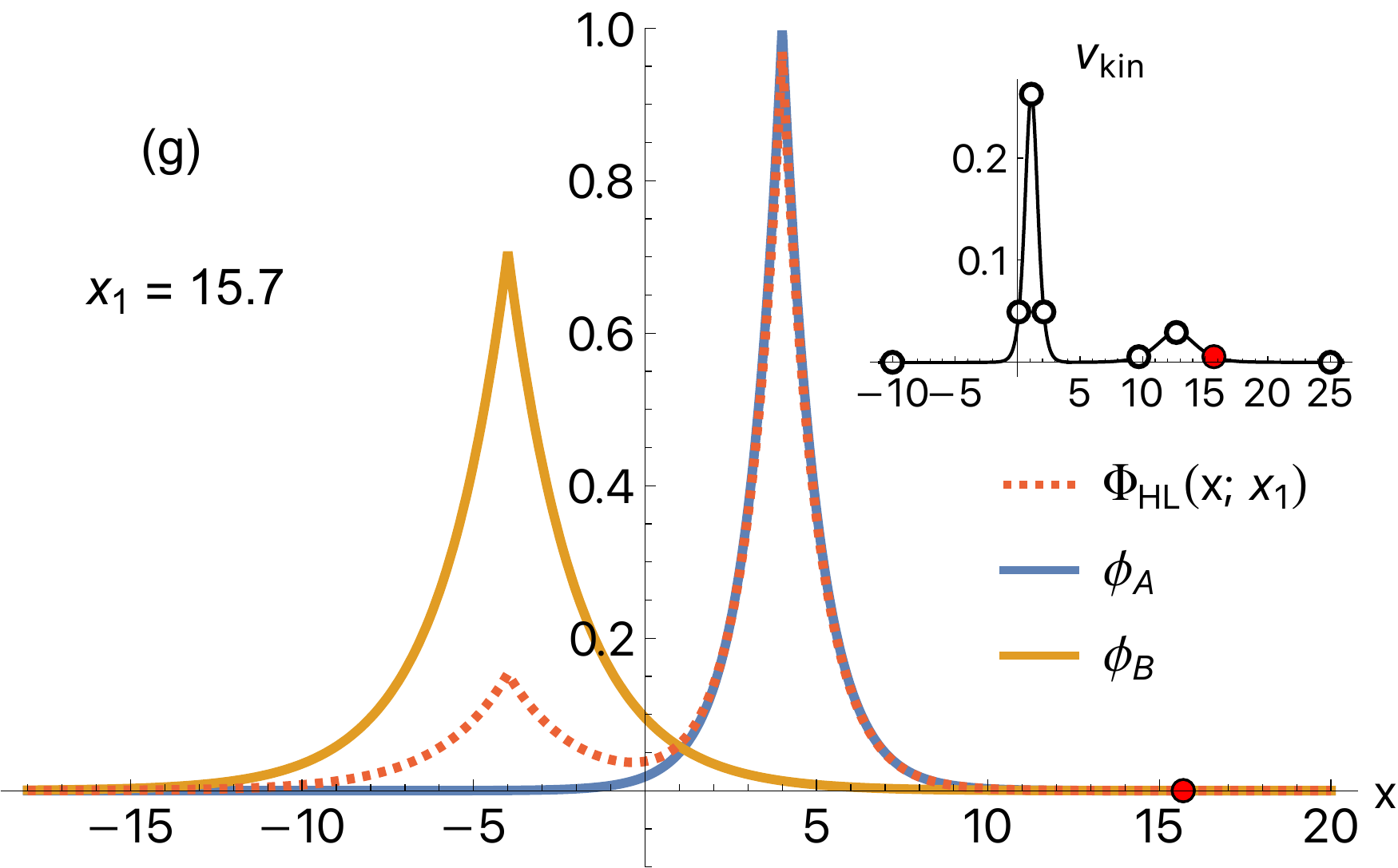}
 \end{subfigure}}
 & {\begin{subfigure}{0.5\textwidth}
 \includegraphics[scale=0.44]{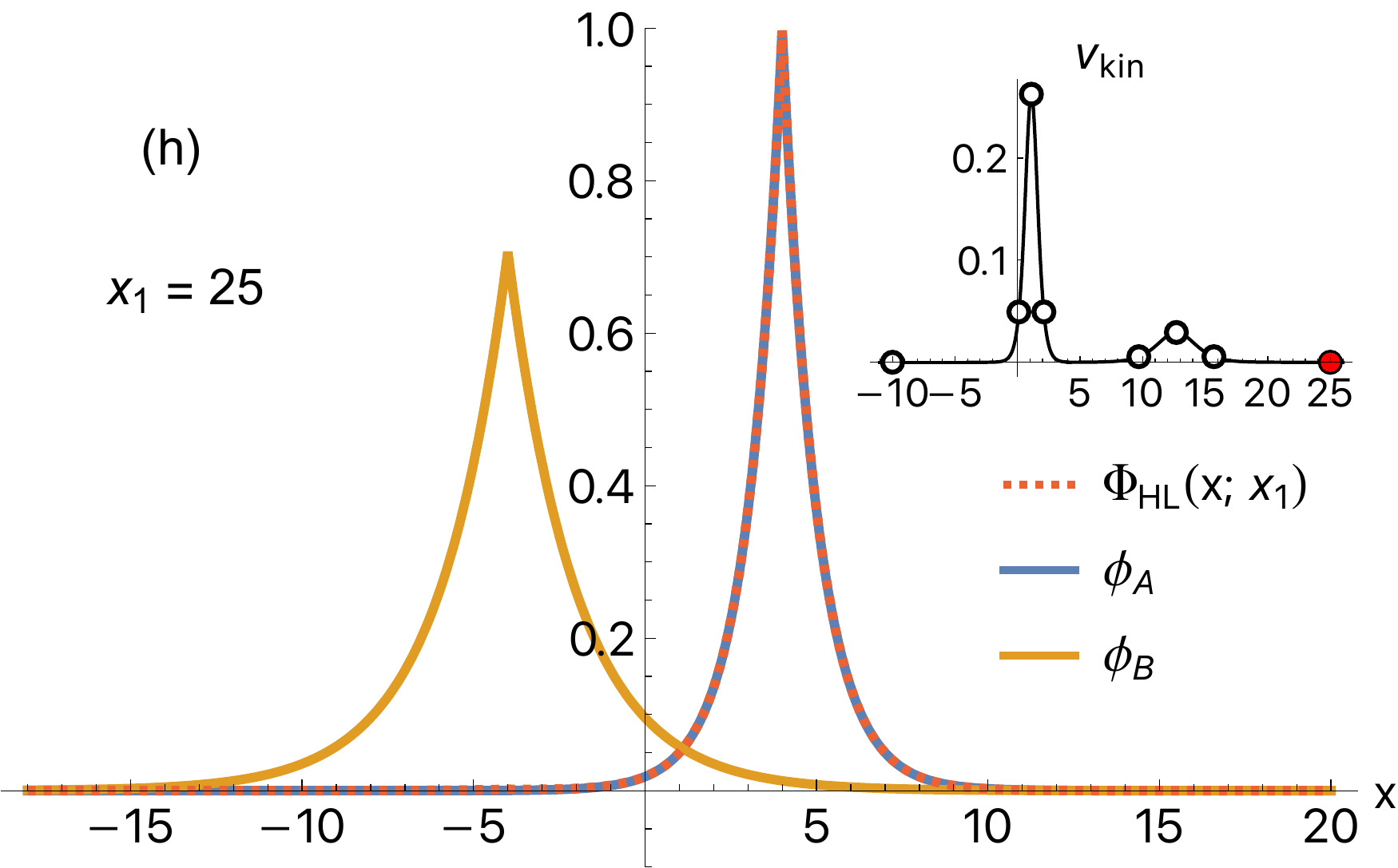}
 \end{subfigure}}
 \end{tabular}   
 \caption{\label{fig:CAjump} Plot of the Heitler--London conditional amplitude, $\Phi_{\HL}(x;x_1)$ (red dashed curve) as function of $x$, for different values of the position $x_1$ of the reference electron (panels a-h). Atomic orbitals $\phi_A(x)$ (blue) and $\phi_B(x)$ (orange) are given by eq~\eqref{eq:expbases} with $a=2$, $b=1$, and $R=8$.  The reference position $x_1$ is indicated on the  horizontal $x$ axis of the main plots with a red circle. The profile of the kinetic potential is shown in the inset on each panel, its value at the reference position being indicated with a red dot.  At very large $\abs{x_1}$ values,  $\Phi_{\HL}(x;x_1)$ resembles $\phi_A(x)$, i.e.\ the one-electron ground state  of the ionised system (panels a and h).
 At the reference positions $x_1=1.1$ and $x_1=12.7$ that correspond to peak positions $x$  of $v_{\kin}(x)$ (panels c and f) the conditional amplitude is approximately a 50-50 mixture of $\phi_A$ and $\phi_B$ (it is in the midst of a jump), whereas at values lying in between the peak values, the conditional amplitude resembles $\phi_B(x)$, i.e.\ an excited ionised state where the only electron is bound by the less attractive nucleus.}
  \end{figure*}

\subsection{The ``jumping" of the conditional amplitude}\label{sec:JoCA}
The definition of the kinetic potential~\eqref{eq:vkin} contains the gradient of the conditional amplitude w.r.t.\ the coordinates of the reference electron. Therefore, we expect this quantity to have maxima when the conditional amplitude is undergoing fast changes. The presence of a secondary kinetic peak at a distance so remote from either nucleus or from the bond midpoint and on the side of the more electronegative atom is somewhat startling. In the following, we provide a clear explanation of the underlying mechanism for this phenomenon.

Let us consider a general two-electron system with two potential wells: nucleus A -- more attractive (to the right) -- and nucleus B -- less attractive (to the left) -- and the two nuclei are `very far' from each other.  By this we mean that the system is effectively dissociated. 
We now look at the character of the conditional amplitude as the reference electron sways along the direction of the stretched bond, on which we focus exclusively. The simple HL model with exponential basis functions is used for the calculation of the conditional amplitude reported in Fig.~\ref{fig:CAjump}.

When the reference electron is located on this axis to the left in a region extending from $ -\infty $ to approximately the bond midpoint, including the B atom region, the conditional amplitude resembles the $1$-electron wavefunction of the A atom. This is the ground state of the ion, and understandably it is also the one-electron wave-function that describes the probability distribution of the remaining electron when the other electron is at the B atom side.  We shall indicate it as $B^+ --- A$.
However, moving on along the axis, the reference electron gradually approaches nucleus A, and over a short distance, depicted in panels b-c-d of Fig.~\ref{fig:CAjump}, the conditional amplitude switches over to the wavefunction of atom B: $B --- A^+$. When the reference electron is in the A atom region, the Coulomb repulsion of the electrons favours the B region for the other electron, which is the first excited state of the ion, with the electron at the less attractive nucleus. This fast change of the conditional amplitude over a short distance causes the kinetic potential to spike. This is similar to the situation for a homonuclear diatomic. When the reference position moves further to the right it leaves the A atom region. At some point the Coulomb repulsion with the reference electron can no longer compete with the stronger attraction of the A nucleus and the electron jumps back to the A atom: $B^+ --- A$. The ion ground state is restored. This is depicted in panels e-f-g of Fig.~\ref{fig:CAjump}. Moving still further to the right, the conditional amplitude remains equal to the ion ground state, which is known to be the limiting shape when the reference position goes to infinity.\cite{KatDav-PNAS-80} The second jump is less sudden, it occurs over a longer range and the corresponding peak of the kinetic potential is much lower than the one in the bond midpoint region. Note that at the two reference positions $x_1=1.1$ and $x_1=12.7$ that correspond to peak positions $x$  of $v_{\kin}(x)$ (panels c and f) the conditional amplitude is approximately a 50-50 mixture of $\phi_A$ and $\phi_B$: we are in the middle of the jump. We conclude that both the primary and the secondary kinetic peak come from the exact same phenomenon: the jumping of the conditional amplitude.

In Ref.~\onlinecite{TemMarMai-JCTC-09}, it was also noted that the first inflection point of the step of the $v_{N-1}$ potential~\eqref{eq:vNm1} for the HL model with exponential bases coincides with $x_{\text{peak}}^{-}$; while in Ref.~\onlinecite{GiaVucGor-JCTC-18} it was shown that both inflection points coincide with $x_{\text{peak}}^{-}$ and $x_{\text{peak}}^{+}$, respectively. In other words the kinetic potential peaks at the borders of the region over which the plateau extends, as assumed in~\eqref{eq:plateauwidth}.
This is now readily understood in terms of the jumping of the conditional amplitude: in the region outside the kinetic peaks the conditional amplitude resembles the ground state wavefunction of the ionised system $B^+ --- A$, i.e.\ $\phi_A$ in our simple model or the antisymmetrized product $\{\Psi_B^{N_B-1}\Psi_A^{N_A}\}$ for a general dissociating $N$-electron system.  The first term on the r.h.s.\ of $v_{N-1}$ in~\eqref{eq:vNm1} is simply the ground state energy of the ionised system $E(B^+ --- A)=E_0^{N-1}= E_A =E_0^N+I_B$. Calling the region inside the kinetic peaks the atom A region, $\Omega_A$, this means $v_{N-1}(\br \notin \Omega_A)=E(B^+ --- A)-E_0^{N-1}=0$. (We will return below to the designation of the region $\Omega_A$.) However, inside $\Omega_A$ (in between the $v_{\kin}$ peaks), the conditional amplitude resembles $B --- A^+$, i.e.\ $\phi_B$ in our simplified model or the antisymmetrized product $\{\Psi_B^{N_B}\Psi_A^{N_A-1}\}$ in general. The first term of $v^{N-1}$ \eqref{eq:vNm1} then corresponds to ionisation of the less electronegative fragment, $E(B --- A^+) = E_0^N+I_A$. This results in $v_{N-1}(\br \in \Omega_A)=E_0^N+I_A-E_0^{N-1}=I_A-I_B$. So the height of the plateau is equal to the difference in ionisation potentials. This was derived with essentially the present argument in Ref.~\onlinecite{GriBae-PRA-96} and was predicted earlier by \citet{AlmBar-INC-85} and Perdew.\cite{Per-INC-85}
  
Our description reveals the link between peaks and plateau structure and shows that the same change of character of the conditional amplitude that is responsible for the primary and secondary kinetic peaks is also responsible
for the rise and the return to zero of the $v_{N-1}$ potential.

\subsection{Relation between crossings of localised orbitals and nodes in delocalised orbitals}\label{subsec:crossings}
The HL model that has been used so far is not the exact wavefunction (except possibly in the limit $R \to \infty$). We have noted that the kinetic peaks in the HL model with exponential basis functions~\eqref{eq:expbases} occur at the points where these latter cross ($\phi_A=\phi_B$). So the locations of the kinetic peaks seem to depend on the choice of the basis. However, it should also be mentioned that with different orbitals, the concurrence of points where the basis functions intercept and of those where the kinetic potential~\eqref{eq:vkinwithoverlap} has its maxima is lost. We consider here the choice of local orbitals in this model in relation to the exact wavefunction. In Sec.~\ref{sec:COMPdet} very accurate wavefunctions will then be considered for a model system.

\begin{figure}[t]
\includegraphics[width=\columnwidth]{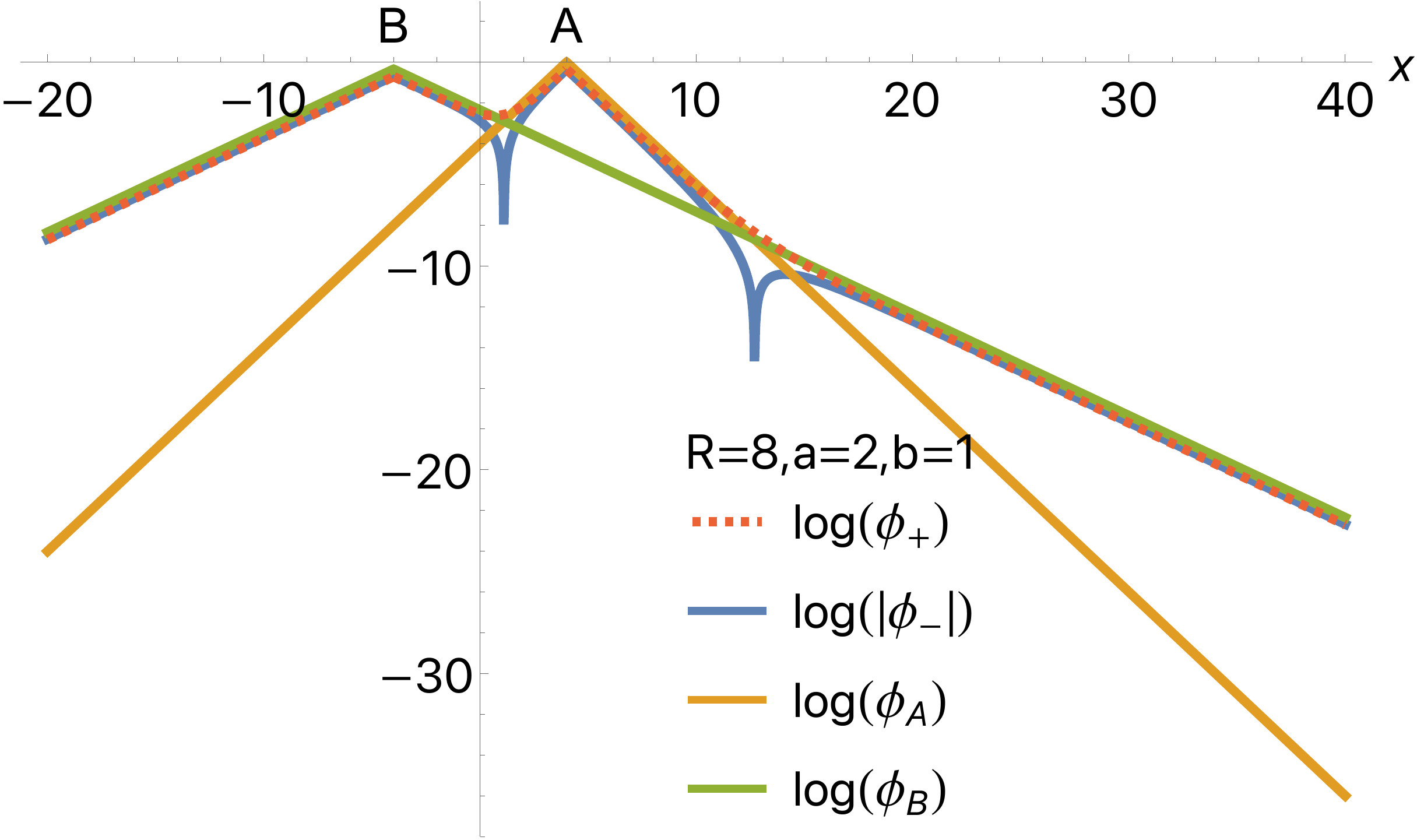}
\caption{The natural logarithm of $\abs{\phi_+}$, $\abs{\phi_-}$, $\phi_A$ and $\phi_B$ are shown. The points at which $\phi_A\equiv \phi_B$ correspond to nodes in $\phi_-$.}
\label{fig:nodes_are_crossings}
\end{figure}

Starting with the given local orbitals [e.g.\ the AOs of~\eqref{eq:expbases}] delocalised orbitals (MOs) can be constructed
\begin{subequations}
\begin{align}
\phi_+ &= \frac{\phi_A+\phi_B}{\sqrt{2(1+S)}} , \\
\phi_- &=\frac{\phi_A-\phi_B}{\sqrt{2(1-S)}} .
\end{align}
\end{subequations}
Note that, if the exponents $a$ and $b$ were equal, $\phi_+$ would correspond to the $\sigma_g$ orbital and $\phi_-$ to the $\sigma_u$. Moreover, while the $\sigma_u$ has only one node exactly at the midpoint, $\phi_-$ presents two nodes exactly where $\phi_A $ and $\phi_B$ intercept each other, see in Fig.~\ref{fig:nodes_are_crossings} the singularities in the plot of the logarithm of $\phi_-$, which occur exactly at the crossings of the (logarithms of) the basis functions. We refer to the region between these singularities (visible in Fig.~\ref{fig:nodes_are_crossings}) as the atom A region $\Omega_A$ invoked in the previous section, reiterating that it nearly coincides with the region inside the kinetic peaks. In Sec.~\ref{sec:Discussion}, we shall give a more physically grounded definition of $\Omega_A$.

Now consider the Slater determinants obtained from doubly occupying these MOs. Their spatial part reads
\begin{subequations}
\begin{align}
\Psi_+ (x_1, x_2)&= \phi_+(x_1)\,\phi_+(x_2) , \\
\Psi_- (x_1, x_2)&= \phi_-(x_1)\,\phi_-(x_2).
\end{align}
\end{subequations}
A configuration interaction wavefunction can be built from the above basis as
\begin{equation}
\Psi_{D} = c_1\Psi_+ + c_2 \Psi_-,
\end{equation}
where we have used the subscript $D$ to indicate that only Slater determinants with doubly occupied orbitals are used and $c_1^2 + c_2^2 = 1$ for normalization.

In principle $c_1$ and $c_2$ are to be determined by minimising the energy corresponding to the expectation value of the total Hamiltonian on $\Psi_D$. However, choosing
\begin{subequations}
\begin{align}
c_1 &= \frac{1 + S}{\sqrt{2(1+S^2)}} , \\
c_2 &= -\frac{1 - S}{\sqrt{2(1+S^2)}}
\end{align}
\end{subequations}
we obtain the HL wavefunction.
If the AOs would be chosen as the solutions of the separate atomic problems, the HL wavefunction would become exact at infinite bond distance.  At finite bond distances the choice of the shape of the localized orbitals in the HL \textit{Ansatz} needs not be dictated by the limiting form of the AOs of the separate atomic problems. The most natural approach to determine them is to use the variational principle to select the optimal ones. This can indeed be implemented, but since this is a two-electron system, we can extract the optimal local orbitals directly from the exact wavefunction. The exact wavefunction can be expanded in a complete set of products of  one-electron basis functions $\{\chi_i\}$, which can be diagonalized\cite{LowdinShull1956, PhD-Giesbertz2010}
\begin{align}\label{eq:psiDiag}
\Psi(x_1,x_2) &=\sum_{i,j} C_{ij}\chi_i(x_1)\chi_j(x_2) \notag \\
&= \sum_ic_i\psi_i(x_1)\psi_i(x_2) ,
\end{align}
where $\psi_i(x)$ are natural orbitals (NOs) and the natural amplitudes [``eigenvalues of the wavefunction'' in the two-electron case] are related to the natural occupation numbers [eigenvalues of the spin-integrated one-body reduced density matrix (1RDM)] as $2\abs{c_i}^2 = n_i$. A compact approximation to the wavefunction can be obtained by truncating the expansion~\eqref{eq:psiDiag}. One can show that the optimal%
\footnote{Optimal is here in the $L^2$ norm, i.e.\ minimal $\norm{\Psi_{\text{exact}} - \Psi_{\text{approx}}}^2$.} 
approximate wavefunction is identical to the diagonal form in~\eqref{eq:psiDiag} in which only the highest occupied NOs are included in the sum and the coefficients are renormalized.\cite{Lowdin1960, Giesbertz2014} Note that this is a special property of the NO expansion for a two-electron system and does not hold in general.\cite{BytautasIvanicRuedenberg2003, Giesbertz2014}

In the dissociation limit, all coefficients tend to zero except the first two. Hence, only retaining the first two configurations gives a good description of the exact wavefunction in this limit. This approximation containing only the first two NOs can readily be transformed at any distance into a HL form~\eqref{eq:HLWF} by constructing the following two non-orthogonal, but normalized localized orbitals
\begin{align}\label{eq:localNOs}
\phi_{A/B} = \frac{\sqrt{\abs{c_1}}\psi_1(x) \pm \sqrt{\abs{c_2}}\psi_2(x)}{\sqrt{\abs{c_1} + \abs{c_2}}} .
\end{align}
It depends on the phase of the NOs whether the plus combination will localize at fragment $A$ or $B$ and \textit{vice versa} for the minus combination. The overlap is readily determined as
\begin{equation}
S = \braket{\phi_A}{\phi_B} = \frac{\abs{c_1} - \abs{c_2}}{\abs{c_1} + \abs{c_2}} .
\end{equation}
It may be  expected that these localised NOs will constitute optimal HL AOs at long but finite distances. We will investigate to what extent using them will change the locations of the crossings depicted in Fig.~\ref{fig:nodes_are_crossings}. It is expected that again the second NO, $\psi_2(x)$ will, like $\phi_-$ above, exhibit nodes at these crossings. A 1D model can afford very accurate solutions of the wavefunction from which the present picture with crossings of localised orbitals  (the HL AOs) and nodes in the delocalised orbitals (either $\phi^-$ or the second NO) can be verified. This is the subject of the next section.\\

\section{Assessment of the Heitler--London \textit{Ansatz}}\label{sec:COMPdet}
The features of the KS potential shown in Figs \ref{fig:resp_het} and \ref{fig:2nd_kin_peak_het} were obtained from accurate solution of a 1D Schrödinger equation (see Sec.~\ref{sec:modelHam}). They have until now been explained and interpreted with a simple HL model wavefunction.  In this section we will verify that the HL model is adequate and does not lead to spurious results and explanations.

 \subsection{Model Hamiltonian}\label{sec:modelHam}
 Figs \ref{fig:resp_het} and \ref{fig:2nd_kin_peak_het} were based on ``exact'' (highly accurate) solutions of the following Hamiltonian
\begin{align}\label{eq:genHam}
\hat{H} =
{}& - \frac{1}{2} \left( \frac{\partial^2}{\partial x_1^2} + \frac{\partial^2}{\partial x_2^2} \right) \notag \\
&{}+ v_{ee}(x_1 -x_2) + v(x_1) + v(x_2) ,
\end{align}
with
\begin{equation}\label{eq:SC}
v_{ee}(x)=\frac{1}{\sqrt{\gamma+x^2}}
\end{equation}
and
\begin{equation}\label{eq:extpot}
v(x) = - \frac{1}{\sqrt{\alpha + \left( x - \frac{R}{2}\right)^2}}-\frac{1}{\sqrt{\beta + \left( x + \frac{R}{2}\right)^2}}.
\end{equation} 
The potential $v_{ee}$ is repulsive, while $v$ is made-up of two potential wells representing the (different) nuclei, separated by a distance $R$.
Both functions are examples of soft-Coulomb interaction, which is one possible choice of interaction function that mimics, in 1D, the effect of three-dimensional Coulomb interaction among charged particles. In 1D, the singularity present in the Coulomb function would lead to spurious nodes in the wavefunction; this singularity is removed by the presence of a softening parameter at the denominator.
The softening parameters, $\alpha$, $\beta$, and $\gamma$ determine the depth (or height) of the potential well (or bulge). Generally speaking, a larger softening parameter results in a shallower potential.
We have chosen $\alpha=0.7$, $\beta=2.25$  and $\gamma=0.6$. These values have no special physical meaning \textit{per se}; however this combination results in a difference in the ionisation potentials of the isolated fragments, $\Delta I$, which is $0.3\,E_h$, reproducing the $\Delta I$ of the real LiH molecule and has been previously adopted in Ref.~\onlinecite{TemMarMai-JCTC-09}.

We have so far explained the results obtained with this Hamiltonian using a simple HL model wavefunction with exponential basis functions~\eqref{eq:expbases}. The HL model is not exact at finite bond distances. Moreover, the exact solutions of the soft-Coulomb attraction are not the exponential basis functions that were used in the HL model, although having exponential decay. We shall therefore address the question of how well this model can capture the physics of Hamiltonian~\eqref{eq:genHam} in the next section.

Details on the precision of the numerical methods used for solving the Schrödinger equation with the Hamiltonian \eqref{eq:genHam} can be found in Appendix~\ref{sec:numdetails}. In the following subsection (\ref{sec:Results}) the comparison of the (numerically) exact solutions with various HL models will be discussed.

\subsection{Numerical results}\label{sec:Results}

In Fig.~\ref{fig:2nd_kin_peak_het}, we show the (correlation) kinetic potential~\eqref{eq:vkin} corresponding to the exact ground state wavefunction of our model Hamiltonian at different internuclear distances. 
The secondary peak is clearly visible despite being an order of magnitude smaller than the primary one. Both primary and secondary peaks increase with increasing $R$ but the increase levels off. So these features of the KS potential change as the geometry changes but they acquire a definite asymptotic shape after a critical distance, showing a saturation behaviour.\cite{GriBae-PRA-96, TemMarMai-JCTC-09, YinBroLopVarGorLor-PRB-16}
A similar saturation behaviour is observed for the response potential, as shown in Fig.~\ref{fig:resp_het}. 
This potential builds a plateau that ``quickly" reaches the asymptotic height of $0.3 \,E_h$, i.e.\ the difference in the ionisation potentials of the separated fragments.
Comparison of the two figures shows that,  for larger bond  distances, the kinetic peaks are located at the borders of the plateau. 
In particular, we find numerically that, as the internuclear distance increases, the maxima of the kinetic potential get closer to the inflection points of the response potential.  This is in agreement with what was found for the first kinetic peak in Ref.~\onlinecite{TemMarMai-JCTC-09} and for both peaks in Ref.~\onlinecite{GiaVucGor-JCTC-18} for the HL model with simple exponential bases treated in Sec.~\ref{sec:HLreview}.

With these exact results available, we can investigate to what extent the HL model is accurate at finite but large $R$ (we take $R=$\SI{11}{\bohr} as representative for the long distances). 
The exact orbitals of the isolated fragments using the model Hamiltonian are not simple exponential functions. So apart from the exponential functions used until now, some sensible choices for the localised orbitals $\phi_A$, $\phi_B$ to be used in the HL expression are:
\begin{enumerate}
\item  use the ground states of the independent fragments;
\item use the lowest two eigenstates of the ionized system.
\end{enumerate}
Since the soft-Coulomb interaction is long ranged, 1.\ and 2.\ differ significantly.  In fact, option 2 is stabilized w.r.t.\ option 1 by an amount which matches almost quantitatively a Coulomb attraction, $-1/R$, exerted by the additional nucleus. For example, at $R=11$, we have $\epsilon_B =-0.475 $ 
and $\epsilon_A= -0.776$ 
in the first case, and  $\epsilon_B =-0.568 $ 
and $\epsilon_A= -0.867$ 
in the second case. Note that the difference in ionization energy $\left( \epsilon_B-\epsilon_A \right)$ is in turn almost constant.

\begin{figure}[t]
\includegraphics[width=\columnwidth]{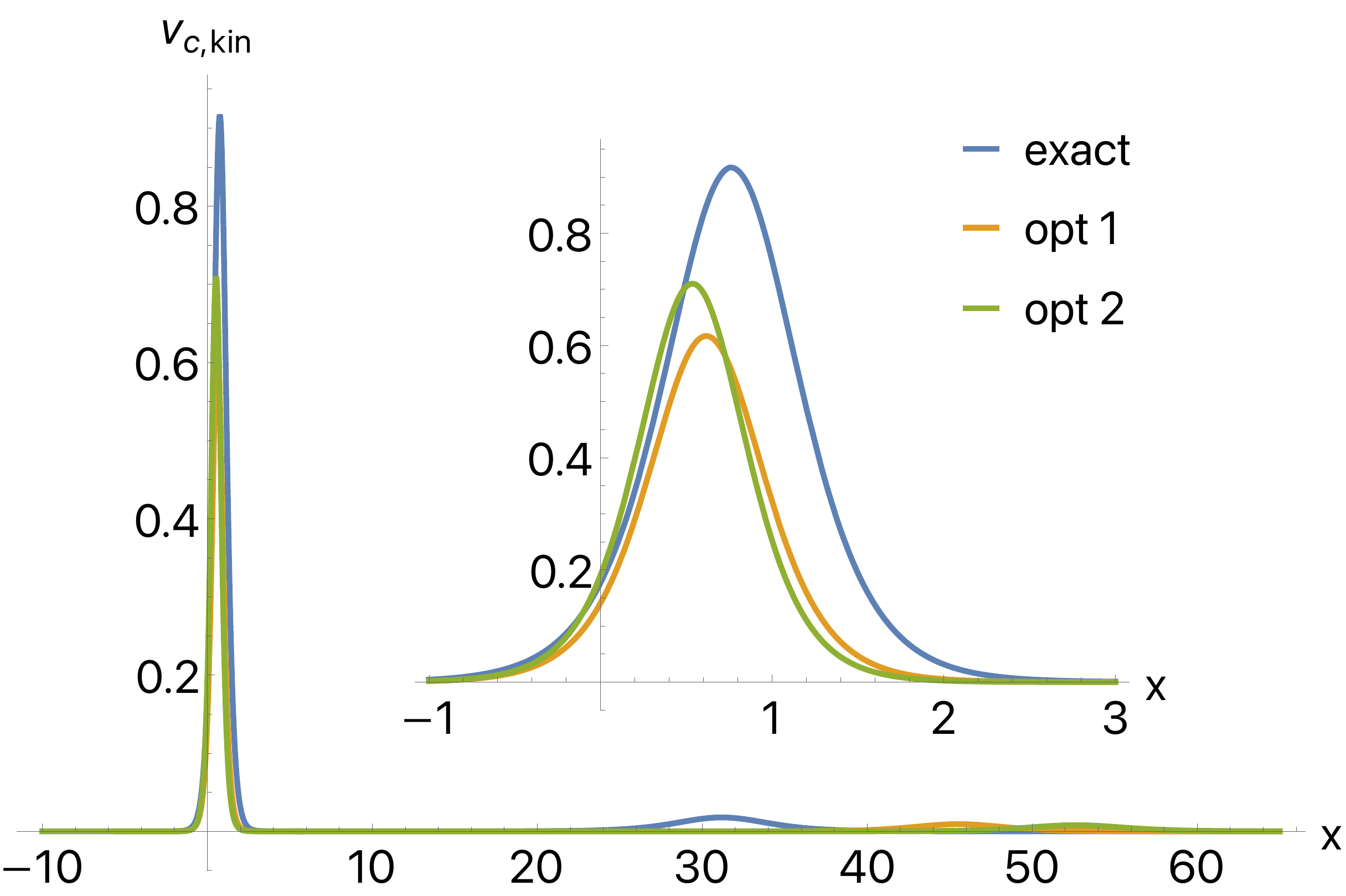}
\includegraphics[width=0.9\columnwidth]{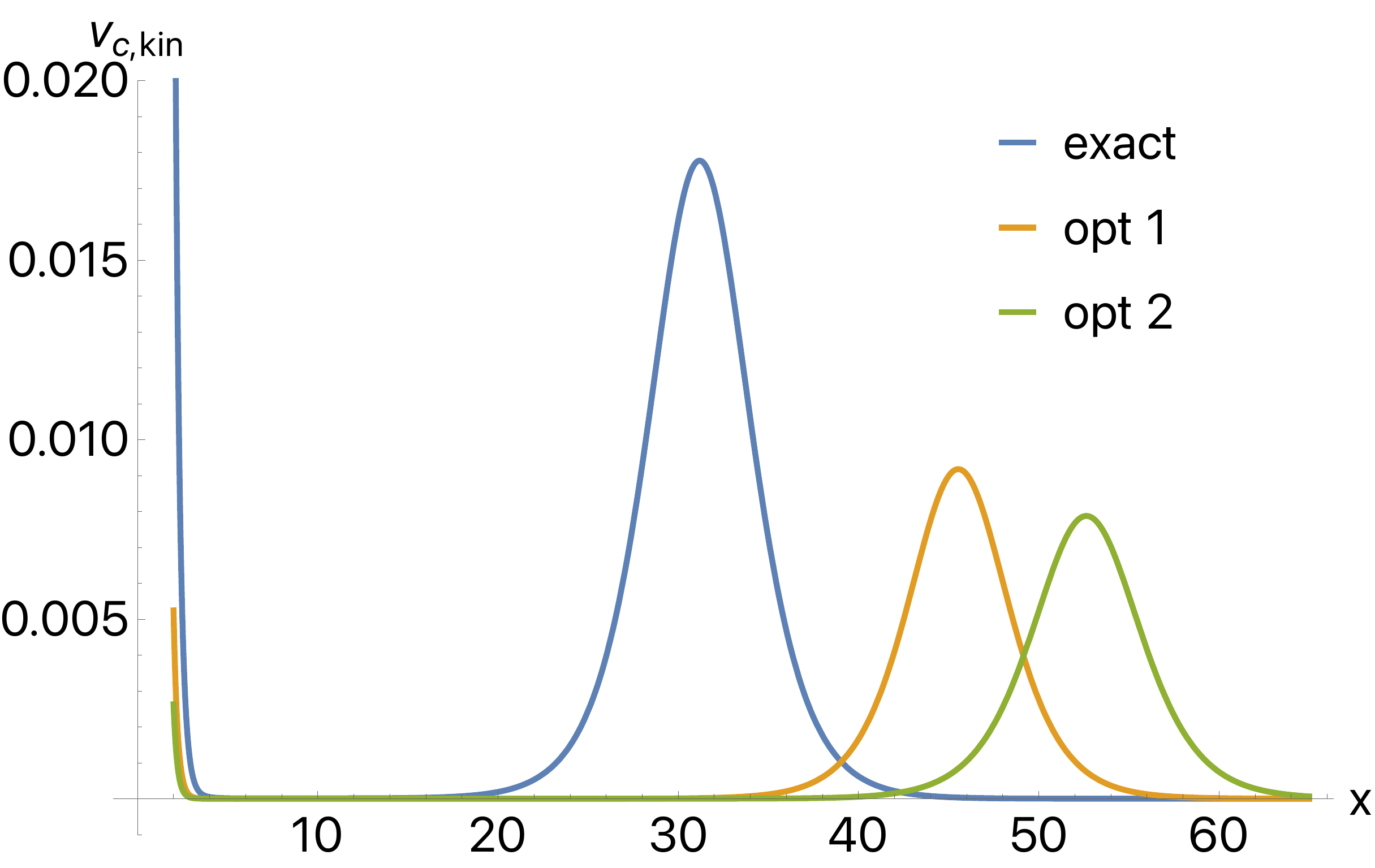}
\caption{Kinetic potential  as obtained from $v_{\kin}^{\HL}$~\eqref{eq:vkinwithoverlap} and exponential basis functions with exponents coming from the calculated ionisation potentials of the separate fragments (orange, option 1) and from the two lowest eigenvalues of the fully ionised system (green, option 2) in comparison with the exact one (blue) for $R=11$.}
\label{fig:vkinfromexponents}
\end{figure}

We first consider two types of simple exponential basis functions~\eqref{eq:expbases} in the HL model, with exponents corresponding to the ionization energies obtained with our model Hamiltonian. So in the equations for $x_{\text{peak}}^\pm$~\eqref{eq:xeq} and $v_{\kin}(x_{\text{peak}}^\pm)$~\eqref{eq:vkinmax}, which are for exponential basis functions, we use the two values for $a = 2\sqrt{2\,I_A}$ with $I_A = -\epsilon_A $ according to cases 1 and 2 above, and the corresponding two values for $b$. This yields HL estimates for both the values at and the locations of the maxima, see option 1 and option 2 for exponential basis functions in Tab.~\ref{tab:vkinestimate}. These estimates are quite off from the exact results, which is also clearly visible in Fig.~\ref{fig:vkinfromexponents}, where we compare the exact kinetic potential (blue) with the HL estimates as obtained from $v_{\kin}^{\HL}$~\eqref{eq:vkinwithoverlap} with the two types of exponential basis functions.

\begin{figure}[t]
\includegraphics[width=\columnwidth]{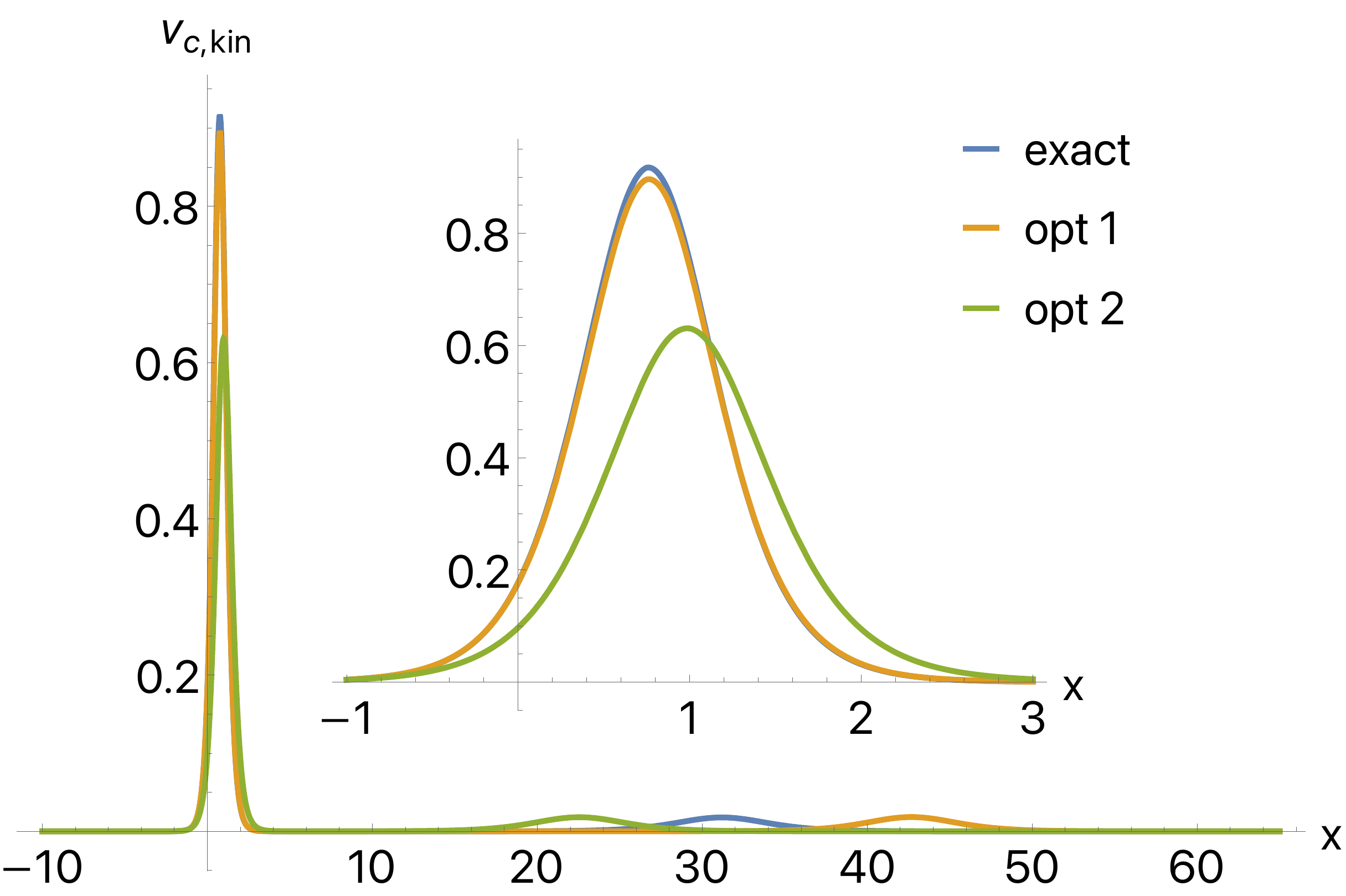}
\includegraphics[width=0.9\columnwidth]{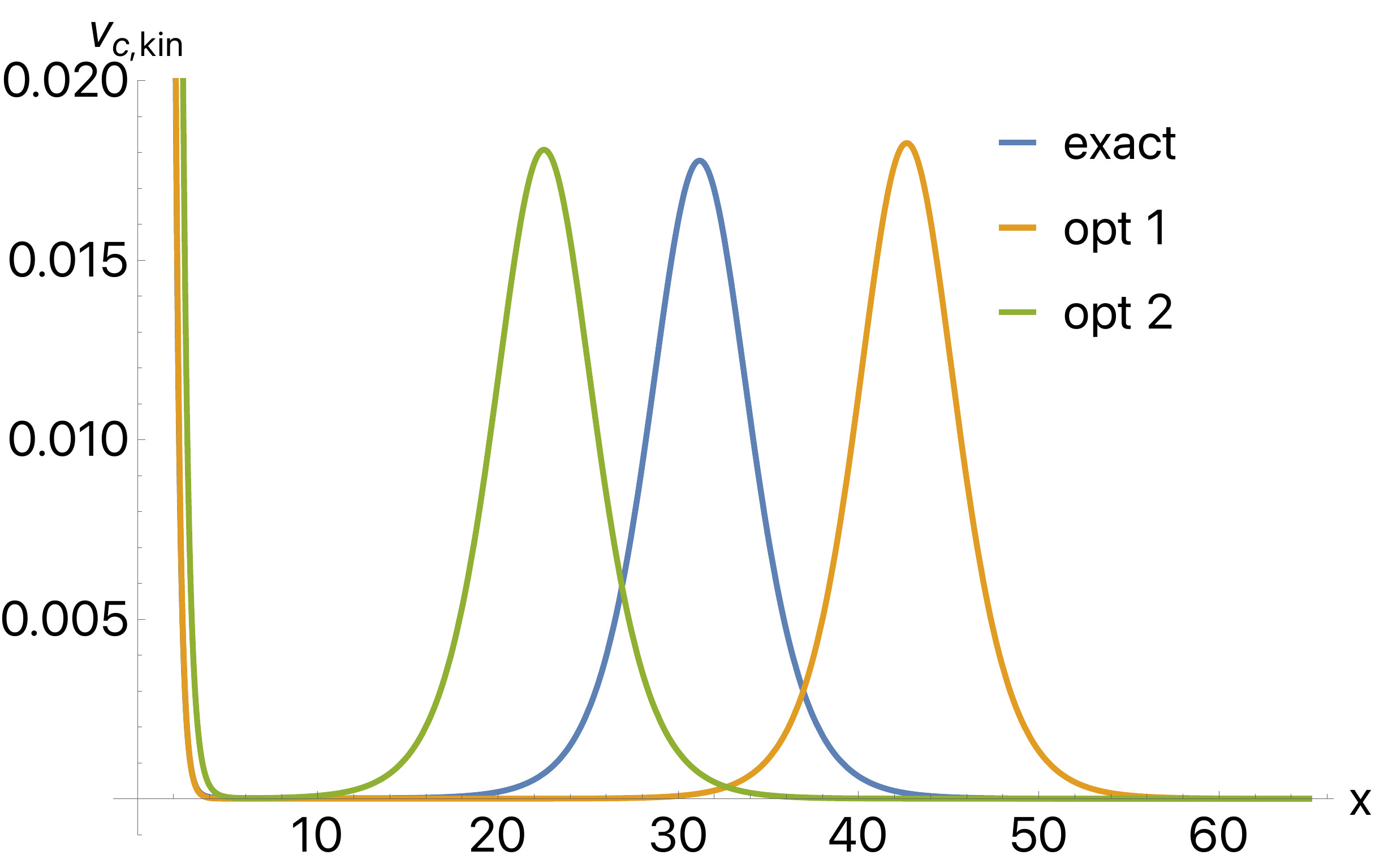}
\caption{Kinetic potential as obtained from forming the HL wavefunction [eq~\eqref{eq:HLWF}] with the separate fragment orbitals (orange) or the two lowest eigenstates of the fully ionised system (green), compared to the exact one (blue) for $R=11$.}
\label{fig:vkinfromorbitals}
\end{figure}

Next, we investigate if an improvement results from choosing as AOs in the HL model the true ``atomic'' solutions of the model Hamiltonian. The kinetic potentials resulting from these two HL wavefunctions are presented in Fig.~\ref{fig:vkinfromorbitals} and in Tab.~\ref{tab:vkinestimate} (option 1 and 2 of ``calculated orbitals''). Clearly, using these calculated AOs gives a much more accurate prediction of the intensity of the peaks and, surprisingly, a change in the ordering of the location of the peaks: now using the two lowest eigenstates of the fully ionised system (green curve) predicts the secondary peak location at shorter distance compared to the exact one, while the use of separate fragment orbitals (orange curve) predicts it at longer distance.

These results indicate that the bare exponential basis is not quantitatively a good model. But they do not rule out the possibility that the HL form of the wavefunction with appropriate localised orbitals could still yield an accurate model for the exact wavefunction at the present long (but finite) distance.  After all, using directly the two types of calculated localised orbitals (options 1 and 2), the accuracy in the intensity of the kinetic peaks has significantly improved and
the estimate obtained with the two different types of localised orbitals now ``sandwiches" the exact location of the peak, hinting at the possibility that some optimal orbitals might work just fine in the HL model.

\begin{figure}[h]
\includegraphics[width=\columnwidth]{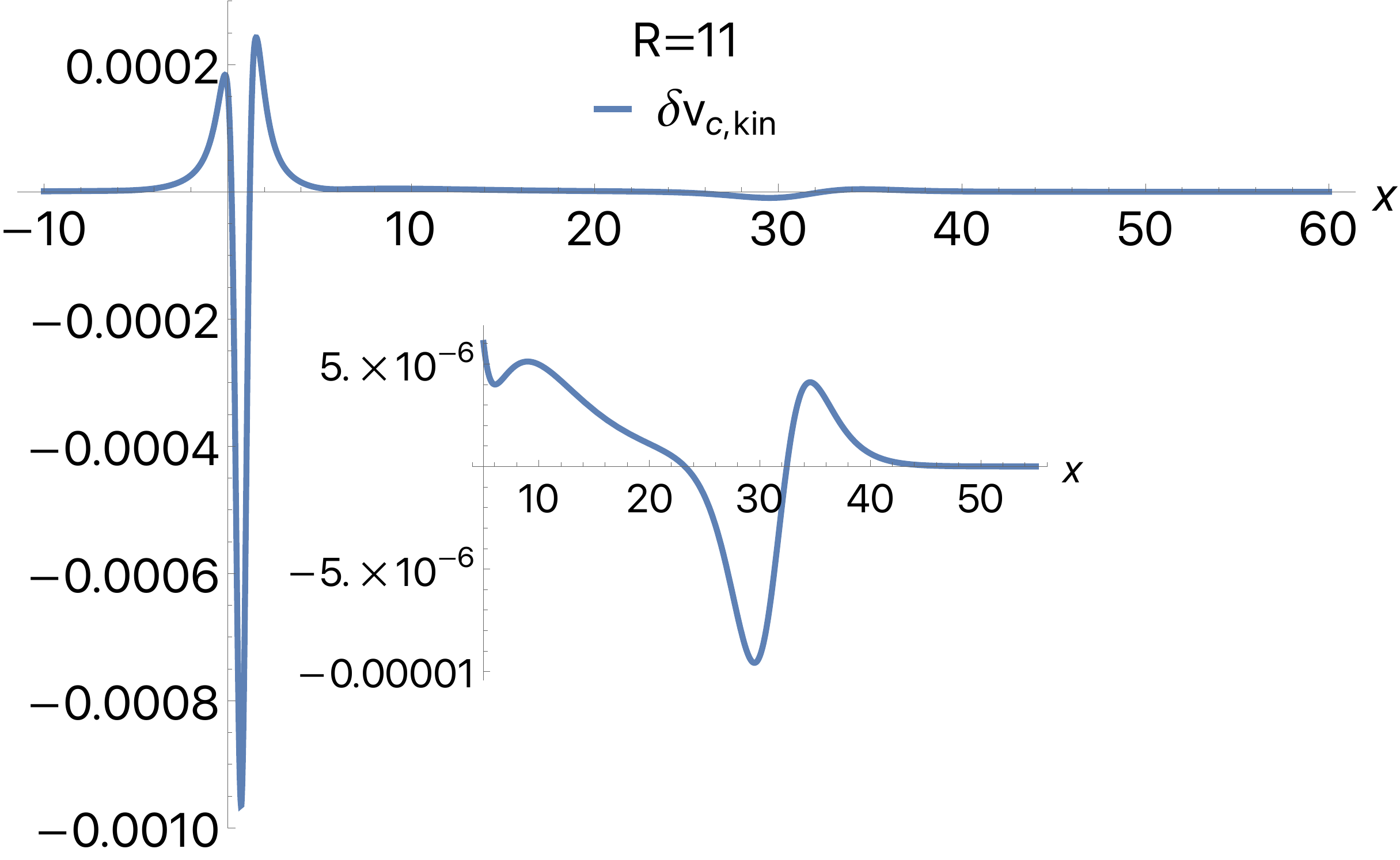}
\caption{Difference, $\delta v_{\cor, \kin}$, between the exact kinetic potential and the one obtained from forming the HL wavefunction [eq~\eqref{eq:HLWF}] with the localised  NOs [eq~\eqref{eq:localNOs}] for $R=11$. In the inset, $\delta v_{\cor, \kin}$ around the secondary peak is enhanced. }
\label{fig:vkinfromNOs}
\end{figure}

This is indeed the case if we choose as localised orbitals linear combinations of the natural orbitals (NOs), see~\eqref{eq:localNOs}. The HL wavefunction thus constructed is an excellent approximation of the exact one and, consequently, it gives an excellent approximation to the kinetic potential. The difference between the exact and the approximate kinetic potential thus obtained is, at most, order $10^{-3}$ as shown in Fig.~\ref{fig:vkinfromNOs}, see also Tab.~\ref{tab:vkinestimate} for the quantitative data on position and height of the peaks.

\begin{table}[b]
\caption{Intensity and location of the kinetic peaks from a HL model wavefunction, according to the different choices for the local orbitals $\phi_A$ and $\phi_B$,  compared to the exact results (last row) for $R=11$ bohr. }
\begin{center}
\begin{ruledtabular}
\begin{tabular}{cccccc}
HL models& &$x_{\text{peak}}^{(-)}$&$x_{\text{peak}}^{(+)}$&$v_\kin(x_{\text{peak}}^{(-)})$&$v_\kin(x_{\text{peak}}^{(+)})$\\
   \hline
   \multirow{2}{*}{$\begin{matrix}
   \text{exponential}\\
   \text{basis}
   \end{matrix}$}&opt 1.&0.62&45.54&0.617&0.009\\
    &opt 2.&0.53&52.64&0.710&0.008\\
    \multirow{3}{*}{$\begin{matrix}
     \text{calculated}\\
   \text{orbitals}
   \end{matrix}$}&opt 1.&0.76&42.68&0.896&0.018\\
    &opt 2.&0.99&22.55&0.630&0.018\\
    &local.\ NOs &0.76&31.19&0.917&0.018\\
    \hline
  exact& &0.76&31.19&0.917&0.018
\end{tabular}
\end{ruledtabular}
\end{center}
\label{tab:vkinestimate}
\end{table}

\begin{figure}[t]
\includegraphics[width=\columnwidth]{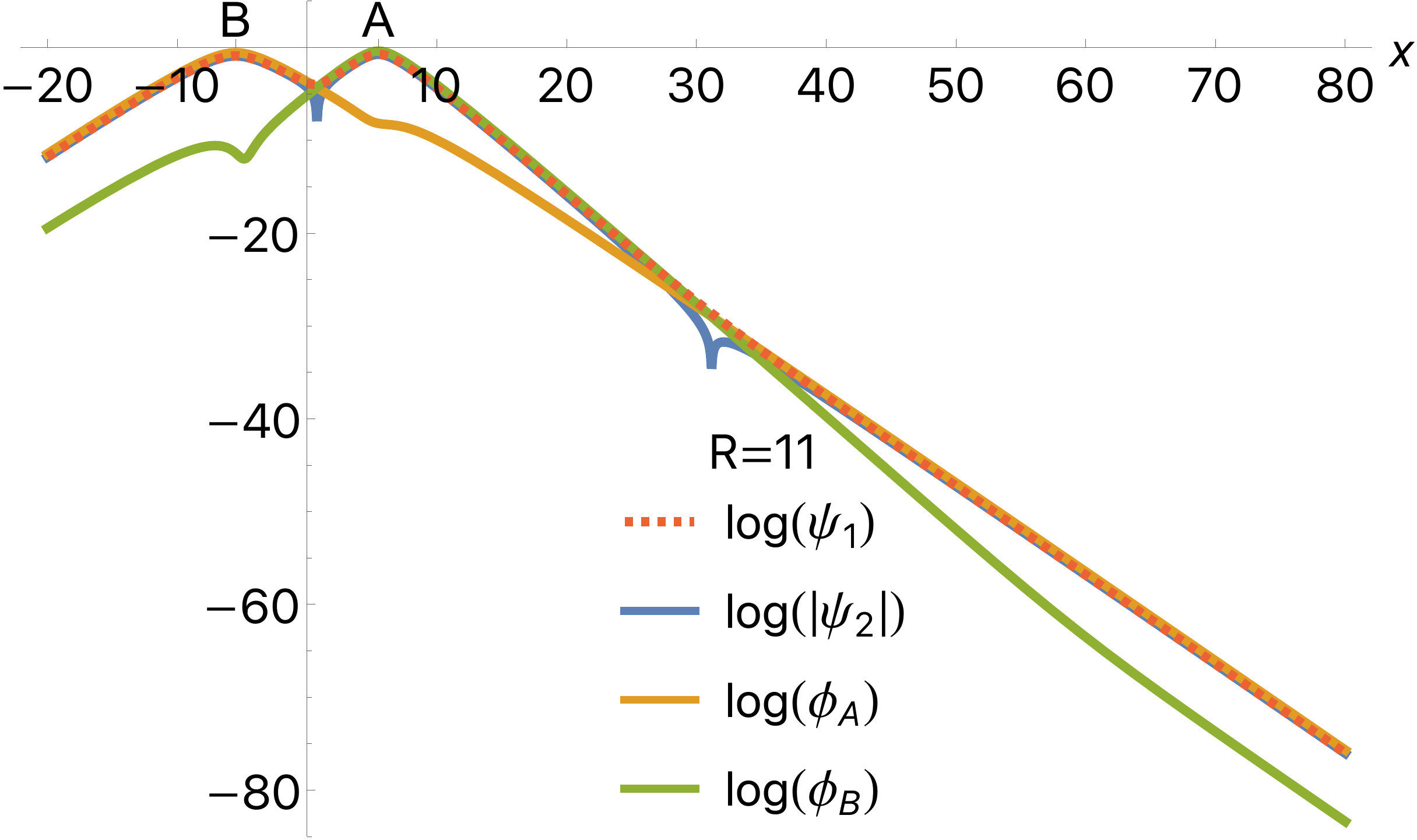}
\caption{The natural logarithm of the natural orbitals in their delocalised and localised (via eq~\eqref{eq:localNOs}) forms. Again, the nodes in the second NO, $\psi_2$ (blue curve), corresponds to crossings of the localised orbitals.}
\label{fig:logNOs}
\end{figure}

That the special features in the KS potential are a genuine effect, connected to the exact wavefunction, is demonstrated in Fig.~\ref{fig:logNOs}, where it is shown that the second NO (see blue curve) has two nodes. Apart from the node close to the bond midpoint, which is expected from its approximate ``$\sigma_u$'' character, there is also a node at long distance. 
The latter coincides with the crossing of the (logarithms of) the two localised NO combinations and is akin to the crossing of the exponential AOs in the HL model (see Fig.~\ref{fig:nodes_are_crossings}) whose important role we highlighted earlier. These nodes in the second NO delineate the atom A region over which the response potential plateau builds up, and whose extent is characterized by the two  $v_{\kin}$ peaks. We note that the two NOs $\psi_1$ and $\psi_2$ have the same asymptotic decay, as they should. The localised combinations of the NOs should have the same asymptotics, but $\phi_A$ only reaches this asymptotics at an extremely large distance (around 70 bohr!). 

Similarly, the picture of the jumping of the conditional amplitude illustrated in Fig.~\ref{fig:CAjump} is reproduced in all its substantial traits when using the (numerically) exact conditional amplitude corresponding to the Hamiltonian of~\eqref{eq:genHam} in contrast with the ground and the first excited states of the ion and can be found in the Supplemental material.

The important conclusion is that the special features of the KS potential can be explained with a simple HL model with exponential basis functions, being not artefacts of that model. In fact, they are intimately connected with the exact solution of the Schrödinger equation, as exemplified by the nodal structure of the second NO.

\section{The conditional potential and the ``atomic regions"}\label{sec:Discussion}
In this section, we focus the attention on yet another feature of the KS potential visible in its conditional component and we revisit the concepts of atomic regions, density ``decay" and ``ionisation potential" in the context of a stretched molecule.

\begin{figure}[t]
\includegraphics[width=\columnwidth]{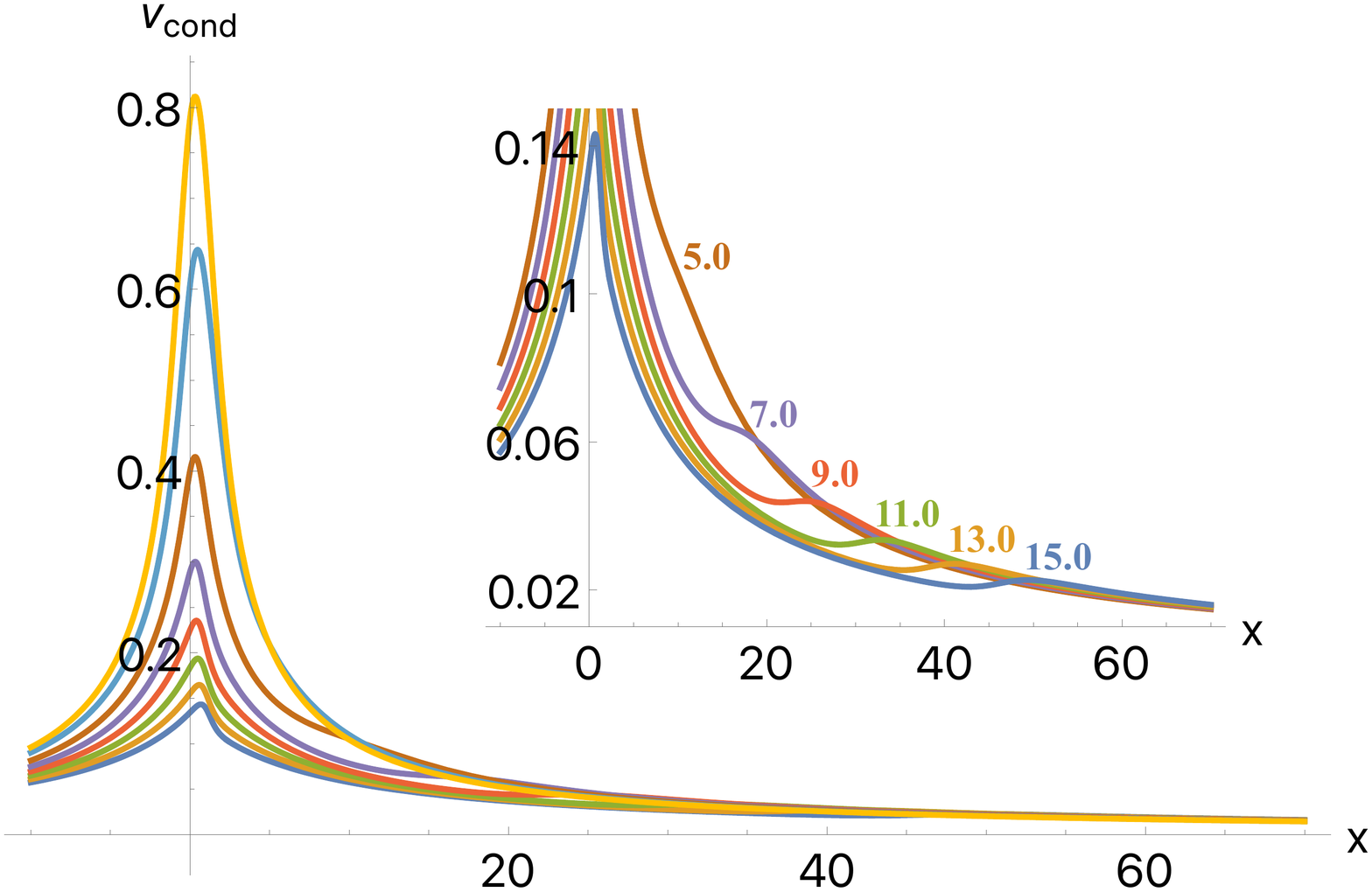}
\caption{Conditional potential, $v_\cond $, at different values of the internuclear distance $R$. In the inset, the increased repulsion between electrons following the jump of the remaining electron around nucleus A is visible.}
\label{fig:condshoulder}
\end{figure}

\begin{figure}[t]
\includegraphics[width=\columnwidth]{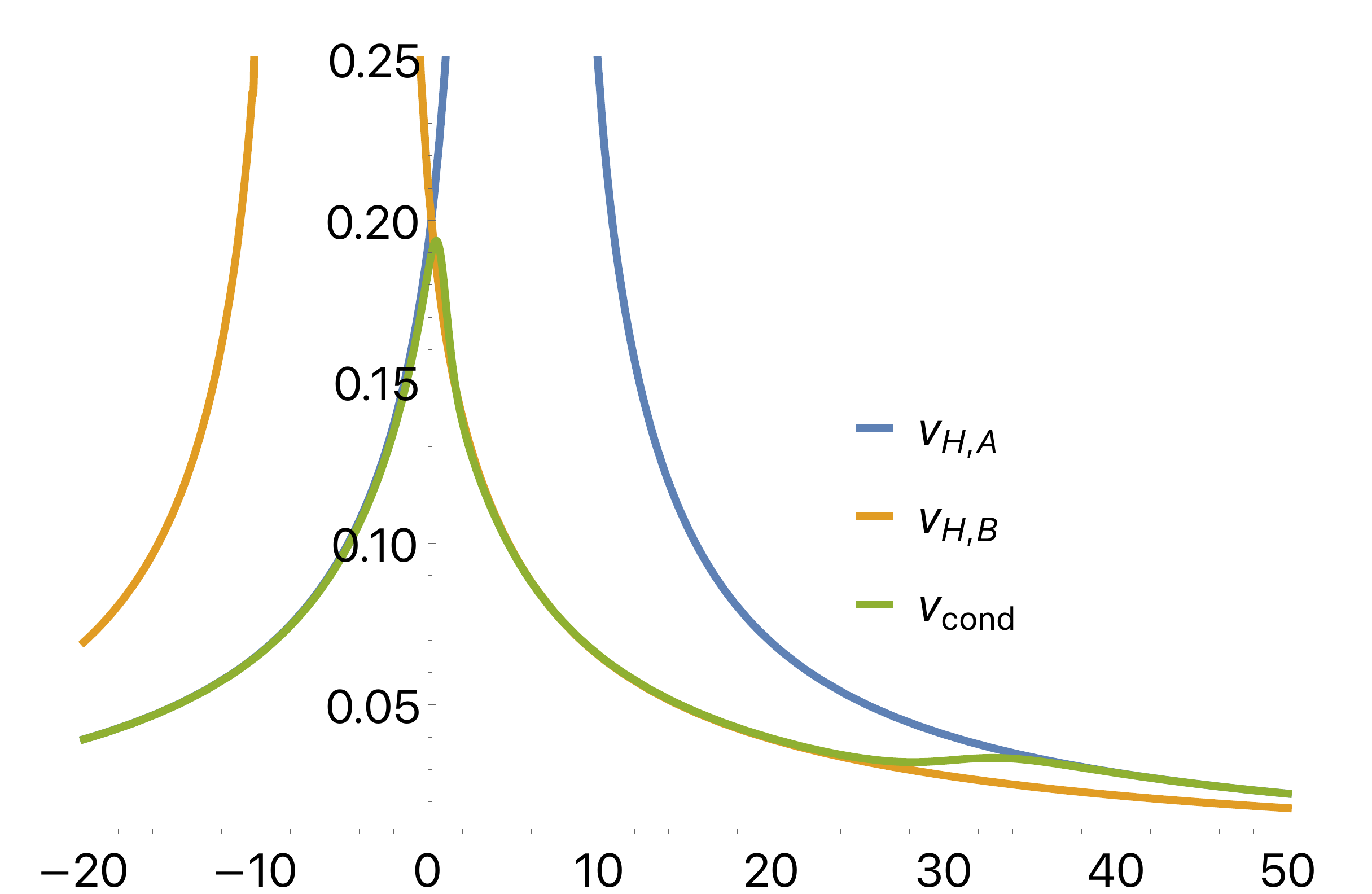}
\caption{Mean-field repulsion from the independent fragments (option 1), $v_{H, A}$ and $v_{H, B}$, in contrast with the conditional potential, $v_\cond$, at $R=11$. In between the fragments, the conditional potential follows closely the profile corresponding to the minimal repulsion, resulting from the conditional amplitude switching from $\phi_A$ to $\phi_B$. The second switch instead is penalised by electron repulsion and resets the conditional potential on the tail of $v_{H, A}$.}
\label{fig:mean-field-pic}
\end{figure}

We stated in Sec.~\ref{sec:JoCA} that moving the reference electron away from atom A to the right, at a certain distance the Coulomb repulsion between the remaining electron and the reference one can no longer compete with the stronger nuclear attraction of nucleus A and the remaining electron jumps back to nucleus A. In addition to observing the effect of this phenomenon from the secondary kinetic peak or from the return to the zero of the response plateau, it can be directly observed from looking at the conditional potential~\eqref{eq:vcond}, plotted in Fig.~\ref{fig:condshoulder} for various internuclear distances. The most striking feature of this plot is the peak close to the bond midpoint, signalling the increased repulsion when the remaining electron gets distributed over the two atomic regions right in the middle of the first jump of the conditional density, see panel (c) of Fig.~\ref{fig:CAjump}, and see also discussion of Fig.~\ref{fig:mean-field-pic}. There is however a second feature far to the right of atom A.
As can be seen from the inset of Fig.~\ref{fig:condshoulder}, at the location at which the remaining electron jumps back around nucleus A the price that is paid in Coulomb repulsion is clearly manifested as a shoulder. 

A decent modelling of the conditional potential at large distances is obtained from the assumption that the two fragments are independent and to first order the reference electron only feels the mean-field repulsion of the electron density in site A, $v_{\Hartree, A}$, if the remaining electron is located around nucleus A, or the mean-field repulsion of the electron density in site B, $v_{\Hartree, B}$,  if the remaining electron is located around nucleus B.

Where the density distribution of the other electron will localise depends, at each position of the reference electron, on the interplay between the mean-field repulsion with each site and the strength of the attraction to each nucleus. In Fig.~\ref{fig:mean-field-pic}, we plot $v_{\Hartree, A}$ and $v_{\Hartree, B}$ as obtained from the independent fragments (option 1) in contrast with the conditional potential, for $R=11$. It is clear that the first jump is strongly driven by electron repulsion: at the peak position, the reference electron has strong repulsion with the remaining electron, whether it is localised on A or B or distributed over the two centers, but when moving further to A the repulsion with the reference electron can be minimised by the conditional amplitude switching from $\phi_A$ to $\phi_B$. On the contrary, the second jump, i.e.\ from $\phi_B$ back to $\phi_A$, is penalised by electron repulsion and plausibly driven by nuclear attraction. Its effect is to reset the conditional potential on the tail of $v_{\Hartree,A}$.

All the features discussed so far hint at the fact that the reference electron experiences a quite isolated environment, insensitive to the presence of the other electron, for the vast majority of the density domain. 
For start, the $v_{N-1}$ potential, which measures the energy of the remaining electron(s) as a parametric function of the position of the reference one is mostly constant. The flat, plateau-like structure of this potential in between the kinetic peaks, indicates that changing the position of the reference electron does not affect the energy of the other one(s). Same goes for the value of this potential outside the kinetic peaks.
After all, the insensitivity of the conditional amplitude to the position of the reference electron is apparent in that the kinetic potential itself is zero for most of the density domain.
In other words, for most of the density domain, if the internuclear distance is large enough, we can regard the two fragments as satisfying two independent Schrödinger equations of the kind
\begin{equation}\label{eq:RSE}
\left(-\frac{\nabla^2}{2} + v_\text{eff} \right) \phi_A = - I_A \phi_A
\end{equation}
with $\phi_A=\sqrt{\ron_A/N_A}\sim e^{-\sqrt{2\,I_A} \, r} $, and similarly for the B site.
In this sense, many authors have used the concepts of ``ionisation potential"  or density ``decay" of the fragments,~\cite{AlmBar-INC-85, Per-INC-85, HodRamGod-PRB-16,HodKraSchGro-JPCL-17,Kraisler2021}
despite the fact that both concepts are strictly defined only for isolated systems and thus, so long as the two fragments are not infinitely apart, there is mathematically only one density decay and one ionisation potential.
The HL model once again helps us making sense of this \textit{conundrum}: because the model is derived in the limit of infinite internuclear separation, it can show asymptotic features at finite distance. In the HL \textit{Ansatz} with exponential fragments, the square root of each fragment density is a solution to the reduced Schrödinger equation with its corresponding Dirac delta potential and the points where the densities of the fragments cross signal the ``borders" of the regions where the reference electron changes from experiencing only one nucleus to experiencing only the other one, as dictated by its exponential behaviour.

In reality, this limiting situation is never attained; however, even in the actual localised NOs coming from our calculations, we could observe a sub-leading exponential behaviour of the NO localised around the more attractive nucleus close to the decay of the independent fragment, which is quite remarkable.
Thus, a sensible definition for the atomic region $\Omega_A$ may be as follows: 
 At large but finite $R$, $\Omega_A$ is the region where the reference electron feels to first order only the A-site field (nucleus A plus a partial screening if $N_A>1$, with $N_A$ the number of electrons bound by this nucleus) and not the B-site field. Analogously, $\Omega_B$ is the region where the reference electron feels only the B-site field.
 
In the limit $R \to \infty$, $\Omega_A$ corresponds more and more with the region in between the crossings of the localised NOs because these localised orbitals are less and less distorted by the presence of the other nucleus. The behaviour of $\phi_A$ at the crossing points is essentially exponential and satisfies to a good approximation the reduced Schrödinger equation for the square root of the density of the A fragment which ties ionisation potential and density decay.
Moreover, with this definition, the fact that $\Omega_B$ extends to the left as well as to the (far) right of nucleus A is easily explained: as the remaining electron jumps from nucleus B to nucleus A, the reference electron at the far right starts to feel nucleus B again -- and stops feeling nucleus A -- as nucleus A is now surrounded by a neutralising electron cloud and, to first order, does not interact with the reference electron.

The atomic regions $\Omega_A$ and $\Omega_B$ do not have clear-cut boundaries: as the reference electron is moving outside of one and into the other, it goes through a transition layer where it feels both nuclei, where indeed the conditional amplitude shows a mixed $\phi_A/\phi_B$ character. This transition may be sharper, as in the primary kinetic peak which is higher and narrower, if the energy gain from transitioning is high (large difference in electron repulsion) or gentler, as in the secondary kinetic peak which is weaker and broader, if the energy gain is small (small difference in nuclear attraction).

The independent fragments picture breaks down locally in these transition layers, where the two fragments show the effects of their interaction by readjusting the electrons on one side of the molecule as a consequence of electron repulsion or nuclear attraction coming from the other side.

 \section{Conclusions}\label{sec:conclusions}
 
In this work, we have performed extremely precise calculations using a 1D model Hamiltonian for a stretched heteronuclear molecule~\eqref{eq:genHam}. The numerical precision accomplished in our code has made it possible to clearly visualise subtle features of the KS potential that have sizeable intensity ($\approx 10^{-3}$ E$_h$) despite appearing in regions where the density is drastically low ($\approx 10^{-40}$ $a_0^{-1}$).
These features show up in specific components of the KS potential known as the response, the kinetic potential and the conditional potential
 (see Figs.~\ref{fig:resp_het}, \ref{fig:2nd_kin_peak_het}, and~\ref{fig:condshoulder} respectively). 
In particular, the kinetic potential~\eqref{eq:vkin} in a stretched heteronuclear molecule shows two peaks: whereas the primary one had been long known, the secondary one has been only recently identified and called for an explanation.
We have provided a description of the mechanism behind the appearance of both primary and secondary peaks of the kinetic potential in terms of the ``jumping" of the conditional amplitude (see Fig.~\ref{fig:CAjump}), elucidating its intimate connection to the response step as well as its effects on the conditional potential.
 
To assess the reliability of the Heitler--London \textit{Ansatz} and of the conclusions that were drawn from it, we have tested different choices of localised orbitals in the HL model and resolved that, if built from optimal orbitals (namely, the localised orbitals obtained as a linear combination of the first two natural orbitals), this model yields an excellent approximation to the (numerically) exact wavefunction at large but finite internuclear distances (see Fig.~\ref{fig:vkinfromNOs}). A remarkable detail emerging from our work is that the second NO presents an extra node on the side of the more electronegative atom very far from the nuclei, at roughly four times the distance of the right nucleus from the origin, see log$(\psi_2)$ in Fig~\ref{fig:logNOs}.
That the second NO has not one but two nodes, underlines the fact that the NOs are not eigenfunctions of a simple Sturm--Liouville equation, i.e.\ a Schrödinger equation with only local (vector) potentials. 
Moreover, its localised combination shows a subleading exponential behaviour very close in value to the decay of the independent fragment bound by the more attractive nucleus, see log$(\phi_A)$ in Fig~\ref{fig:logNOs}. The orbital $\phi_A$ acquires the expected, slower, asymptotic decay extremely far out (well beyond the location of the extra node in the second NO). 

Although all our calculations have been restricted to a two-electron singlet system, a generalisation to any $N$ electron system in which a singlet molecule dissociates into two fragments with one unpaired electron each can be made as follows.
In very low density regions, we can consider that only the HOMO, the most diffuse orbital, will contribute to the density. The remaining orbitals will be mostly concentrated around each nucleus and they will determine the detailed structure of the KS potential for each atom, their ``intra-fragments" structure. (For example steps in the response potential appear when going through the inner shells of atoms and the correlation kinetic potential peaks at the borders of these shells.\cite{GriLeeBae-JCP-94}) 
However, the features that govern the dissociation process, the ``inter-fragments" features, will all be determined by the behaviour of the HOMO alone. In Ref.~\onlinecite{GriBae-PRA-96}, using this very argument it was already concluded that the term $v_{s,N-1}$ in~\eqref{eq:vrespNm1} does not contribute to building the inter-fragment step structure of $v_{\resp}$. 
Similarly, using the definition of the kinetic potential with the KS conditional amplitude, we have
\begin{align}\label{eq:vkin-s-orbs}
v_{s, \kin}(\br)
& = \frac{1}{2}\sum_{i=1}^H\sum_\sigma\abs[\bigg]{\nabla\frac{\psi_i(\bx)}{\sqrt{\ron(\br)}}}^2 \\ 
& = \frac{1}{2\ron(\br)}\sum_{i=1}^H\sum_\sigma\abs{\nabla \psi_i(\bx)}^2- \frac{\abs{\nabla\ron(\br)}^2}{8\ron(\br)^2} , \notag
\end{align}
which clearly vanishes in very low density regions where only $\psi_H$, behaving like the square root of the density, i.e.\ $\psi_H \sim \sqrt{\ron} $, survives.

This means that two-electron singlets are an excellent proxy for studying the breaking of covalent bonds and, most importantly, that we can correctly describe this kind of dissociation processes in any $N$-electron system having an approximation for $v_\kin$ and $v_{N-1}$, whereas $v_{s,\kin}$ and $v_{s,N-1}$ may be disregarded to this end.
Given that the kinetic potential and the $N-1$ part of the response potential have proven the most delicate to model, gaining insight into their structure, as carried out throughout this work, may help the development of approximations for these specific components.

\section{Acknowledgements}
We thank Prof.\ P.\ Gori-Giorgi  for helpful discussions. 
Financial support from the Netherlands Organisation for Scientific Research under Vici Grant No.~724.017.001 is acknowledged by SG and KJHG. SG also acknowledges funding from DOE NNSA grant DE-NA0003866.

\appendix
\section{Numerical details}\label{sec:numdetails}
For all numerical calculations with the model Hamiltonian of~\eqref{eq:genHam} we have used an equidistant grid from 
\SI{-25}{\bohr}
 till \SI{100}{\bohr}.
 For the discretization \num{1000} points per dimension appeared to be sufficient, which amounts to \num{500500} grid points for the full singlet wavefunction. For the second order derivatives we used a regular 10th order finite difference scheme and the Dirichlet boundary conditions were implemented by taking the function anti-symmetric over the border.\cite{NieRugLee-EPJB-18} Only the kinetic energy couples to wavefunction values at other positions (max 20), so the many-body Hamiltonian is very sparse and has been implemented as a subroutine. As a solver we used the filtering technique described in Ref.~\onlinecite{ZhouSaad2007}. Such a Krylov based routine is only suitable to converge the error in the norm $\epsilon_{\text{norm}} = \norm{\hat{H}\Psi - E\Psi}$, but for our asymptotic data we need to converge the relative error $\epsilon_{\text{rel}} = \norm{\hat{H}\Psi/\Psi - E}$. For example, the secondary kinetic peak for $R =$\SI{15}{\bohr} is located in a region where the density is of the order 
 \SI{e-40}{\per\bohr}
  and we need to divide by the square root of this.
To reach this level of precision we switched to only filtering when the Krylov iteration started to deteriorate the eigenvector.\cite{NieRugLee-EPJB-18} With this technique we could easily get the relative error $\epsilon_{\text{rel}}$ down to $\sqrt{\varepsilon} \approx \num{1.5e-8}$, as we worked in double precision. The same routine was also needed to extract the NOs with sufficient accuracy, since the standard LAPACK diagonalization routines only converge the error in the norm and are not built to converge the relative error $\epsilon_{\text{rel}}$ in the eigenvectors.
The code was implemented using the \textsc{Fortran 2018} language specification and parallelized using the \textsc{OpenMP} API. The code is currently only available upon request to provide also the necessary guidance, since proper documentation is currently lacking to make it public.

\bibliography{GiaNeuBaeGie}

\end{document}